\documentclass[reprint,amsmath,amssymb,aps,superscriptaddress,twocolumn,float]{revtex4-2}
\usepackage{graphicx}
\usepackage{graphics}
\usepackage{dcolumn}
\usepackage{bm}
\usepackage{amsfonts}
\usepackage{amssymb}
\usepackage{float}
\usepackage{xcolor}

\begin{document}

\title{Large magneto-optical effect and magnetic anisotropy energy in two-dimensional metallic ferromagnet Fe$_3$GeTe$_2$}

\author{Ming-Chun Jiang}
\affiliation{Department of Physics and Center for Theoretical Physics, National Taiwan University, Taipei 10617, Taiwan\looseness=-1}

\author{Guang-Yu Guo}
\email{gyguo@phys.ntu.edu.tw}
\affiliation{Department of Physics and Center for Theoretical Physics, National Taiwan University, Taipei 10617, Taiwan\looseness=-1}
\affiliation{Physics Division, National Center for Theoretical Sciences, Taipei 10617, Taiwan\looseness=-1}

\date{\today}

\begin{abstract}
The recent discovery of long-range magnetic orders in atomically thin semiconductors
Cr$_2$Ge$_2$Te$_6$ and CrI$_3$ as well as metal Fe$_3$GeTe$_2$ has opened up exciting opportunities
for fundamental physics of two-dimensional (2D) magnetism and also
for technological applications based on 2D magnetic materials.
Due to their unique metallic nature, atomically thin Fe$_3$GeTe$_2$ films
exhibit fascinating properties such as Fermi level-tunable Curie temperature,
magnetic anisotropy energy and anomalous Hall effect, thus offering valuable applications
for, e.g., voltage-controlled 2D spintronics at room temperature.
Nevertheless, to exploit these 2D metallic magnets, the mechanisms that control
their physical properties should be well understood.
In this paper, based on systematic first principle density functional theory calculations,
we study two relativity-induced properties, namely, magnetic anisotropy energy (MAE) and 
magneto-optical (MO) effects, of multilayers [monolayer, bilayer, trilayer,
fourlayer and quintlayer] and bulk Fe$_3$GeTe$_2$ and also their connections
with the underlying electronic structures of the materials.
Firstly, all the considered Fe$_3$GeTe$_2$ structures are found to prefer the out-of-plane magnetization
and have gigantic MAEs of $\sim$3.0 meV/f.u., being about 20 and 6 times larger than 
2D ferromagnetic semiconductors Cr$_2$Ge$_2$Te$_6$ and CrI$_3$, respectively.
This gigantic perpendicular anisotropy results from the large magnetocrystalline anisotropy energy (MCE)
which is ten times larger than the competing magnetic dipolar anisotropy energy (MDE) which always favors 
an in-plane magnetization. The giant MCEs are attributed to the large Te $p_{x,y}$ orbital density
of states near the Fermi level and also to the topological nodal point just below the Fermi level at the K points
in the Brillouin zone.
Secondly, 2D and bulk Fe$_3$GeTe$_2$ also exhibit strong MO effects with their Kerr and Faraday rotation angles
being comparable or even larger than that of best-known MO materials such as PtMnSb, Y$_3$Fe$_5$O$_{12}$ 
and Bi$_3$Fe$_5$O$_{12}$. The features in the Kerr and Faraday rotation spectra
are almost thickness-independent although the Kerr rotation angles increase monotonically with film thickness.
The strong MO Kerr and Faraday effects are found to result from 
the large MO conductivity (or strong magnetic circular dichroism) in these ferromagnetic materials.
In particular, the calculated MO conductivity spectra are one order of magnitude larger than that of Y$_3$Fe$_5$O$_{12}$. 
The calculated MO conductivity spectra are analysed in terms of the symmetry of the band states
and dipole-allowed optical transitions at high symmetry $\Gamma$, K and K$'$ points,
which further reveal that atomically thin Fe$_3$GeTe$_2$ films with odd layer-number 
would exhibit anomalous ferrovalley Hall effect.
All these interesting findings thus suggest that 2D and bulk ferromagnetic Fe$_3$GeTe$_2$ 
are promising materials for high density MO and spintronic nanodevices.      

\end{abstract}

\maketitle

\section{INTRODUCTION} 
Recent material realization of intrinsic magnetism in atomically thin films of semiconducting 
Cr$_2$Ge$_2$Te$_6$~\cite{Gong2017} and CrI$_3$~\cite{Huang2017} as well as metallic 
Fe$_3$GeTe$_2$~\cite{Fei2018,Deng2018} has created numerous fascinating opportunities for two-dimensional (2D) magnetism.
Fundamentally, the famous Mermin-Wagner theorem \cite{Mermin1966} dictates that thermal fluctuation prohibits 
any long-range magnetic order in isotropic 2D systems at any finite temperature. 
The discovery of the long-range ferromagnetic order in monolayers (MLs) CrI$_3$~\cite{Huang2017}
and Fe$_3$GeTe$_2$~\cite{Fei2018,Deng2018} thus demonstrates that theories of 2D magnetism need to go beyond the isotropic Heisenberg model. 
Magnetism at 2D limit is not only highly desirable for the fundamental physics but also for the technological
applications ranging from magnetic memories to sensing, to spintronics to novel functionalities based on 2D materials.
Among the magnetic 2D materials, few-layer Fe$_3$GeTe$_2$ structures are unique since they are apparently the 
only ferromagnetic metal that retains the long-range magnetic order down to the 2D limit~\cite{Burch2018,Gibertini2019},
and thus attract particularly strong attention.
For example, owing to its metallic nature, Curie temperature ($T_C$) of trilayer (TL) Fe$_3$GeTe$_2$ 
was raised from 100 K to 300 K by ionic gating~\cite{Deng2018}, thus offering application potential
for voltage-controlled 2D spintronics at room temperature.  
Furthermore, it was demonstrated theoretically that the magnetocrystalline anisotropy energy (MCE)
of ML Fe$_3$GeTe$_2$ is large and also tunable by either tensile strain~\cite{Zhuang2016} or electric gating~\cite{Wang2020}.
Strong layer-dependent anoamalous Hall effect in atomically thin Fe$_3$GeTe$_2$ was also predicted recently~\cite{Lin2019}.
Interestingly, unlike 2D semiconductors CrGeTe$_3$ and CrI$_3$ which all have a centrosymmetric crystalline structure,
few-layer Fe$_3$GeTe$_2$ with odd number of MLs have the broken inversion symmetry (see Table I below). 
Consequently, atomically thin Fe$_3$GeTe$_2$ with odd layer-number are expected to exhibit
novel properties such as anomalous valley Hall effect~\cite{Tong2016} and magnetically tunable
second-order nonlinear optical responses (e.g., second-harmonic generation and bulk photovoltaic effect)~\cite{Gudelli2020}. 

In this work, we concentrate on two relativity-induced properties of 2D Fe$_3$GeTe$_2$,
namely, magnetic anisotropy energy (MAE) and magneto-optical (MO) effects.
MAE is the energy needed to rotate the magnetization direction 
from the easy axis to the hard axis. MAE plays a crucial role in suppressing thermal fluctuation and
thus stabilizes long-range magnetic orders in 2D systems. It is also an important factor 
that characterizes a magnetic material from the application view point.
In particular, a thin film with a large perpendicular magnetic anisotropy may 
find applications in high density magnetic data storage. 
It has two contributions, namely, the magnetocrystalline anisotropy energy (MCE) and the 
magnetic dipolar anisotropy energy (MDE). MCE is due to the difference 
between the relativistic band structures for two different magnetic orientations. 
On the other hand, MDE originates from the classical magnetic dipole-dipole 
interaction in the magnetic solid~\cite{Guo1991A,Guo2007}. Interestingly, 
MDE is determined solely by the crystalline structure and geometric shape of 
the sample~\cite{Guo1991A,Guo2007}. In a layered material, the MDE always prefers 
an in-plane magnetization while MCE could favor either an in-plane or the out-of-plane 
magnetization~\cite{Yimei2018}.   
Although the MAE of bulk and ML Fe$_3$GeTe$_2$ has been studied 
both experimentally and theoretically~\cite{Zhuang2016,Wang2020,Park2020},
the MAE of atomically thin Fe$_3$GeTe$_2$ films of few-layer thickness has not been reported yet. 

MO effects are manifestations of the interplay between magnetism and light in magnetic 
solids~\cite{MO_Oppeneer_book,MO_Antonov_book}. When a linearly polarized light beam hits 
a magnetic material, the principal axis of the reflected and transmitted light rotates with
respect to the polarization direction of the incident light. The former is called 
the magneto-optical Kerr effect (MOKE) and the latter is known as the magneto-optical Faraday effect (MOFE). 
MO effects originate the simultaneous presence of the relativistic spin-orbit
coupling (SOC) and spontaneous magnetization in the magnetic solid. 
The spontaneous magnetization and the SOC result in energy band splitting and 
thus lead to different refractive indexes for the right- and left-circularly polarized light. 
As will be discussed in the next section, this magnetic circular dichroism (MCD) 
gives rise to the MOKE and MOFE.
MOKE has been widely applied to study the magnetic properties of thin films and surfaces  
\cite{MO_Antonov_book}. Indeed, in Fe$_3$GeTe$_2$ and other 2D ferromagnets, 
long range magnetic orders were discovered by the MOKE and MCD 
experiments~\cite{Deng2018,Fei2018,Huang2017,Gong2017}. Large MOKE and MOFE effects can also be 
exploited for fabricating high density MO data storage and MO sensors.~\cite{MO_storage,MO_device} 
On the other hand, it is known that 3$d$ transition metal alloys that contain heavy elements 
such as FePt would have large MO effects due to the large SOC strength on the heavy element atoms~\cite{Guo1995,Guo1996}.
Since Fe$_3$GeTe$_2$ contains heavy Te atoms, bulk and 2D Fe$_3$GeTe$_2$ are expected to have large MO effects.
However, there has been no theoretical study on the optical and MO properties of bulk and few-layer Fe$_3$GeTe$_2$.

In this work, therefore, we carry out a systematic first principles density functional theory (DFT) study 
on the magnetic, electronic, optical and MO properties of monolayer (ML), bilayer (BL), trilayer (TL),
fourlayer (FL) and quintlayer (QL) as well as bulk Fe$_3$GeTe$_2$.
Indeed, we find that all the considered Fe$_3$GeTe$_2$ structures would exhibit strong MOKE and MOFE
especially in the infrared and visible frequency range. 
Furthermore, we also find that all the structures have a large MAE, being comparable to
that of FePt which has the largest MAE among the ferromagnetic transition metals and their alloys.
These findings suggest that bulk and few-layer Fe$_3$GeTe$_2$ are promising ferromagnetic materials for 
nanoscale MO and spintronic devices. 

The rest of this paper is organized as follows.
In the next section, a brief description of the crystalline structure of
the considered Fe$_3$GeTe$_2$ structure as well as the used theoretical methods
and computational details is given. In Sec. III, the calculated magnetic properties and
electronic band structures as well as the optical and MO properties are presented.
The possible origins of the large MAEs are discussed in terms of orbital-decomposed
density of states (DOSs) and also the topological features of the band structure near the
Fermi level. Also, the peaks in the calculated MO conductivity spectra are
analyzed in terms of the symmetry of the energy bands and main interband optical transitions
at high symmetry $\Gamma$ and K points in the Brillouin zone (BZ). 
Finally, the conclusions drawn from this work are summarized in section IV. 
 
\begin{figure}[tbph] \centering
\includegraphics[width=8.0cm]{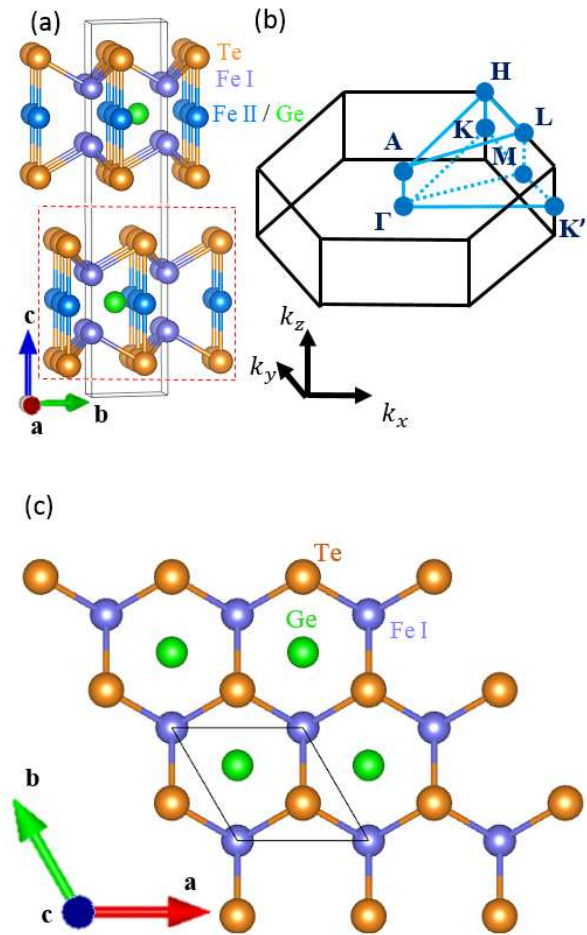}
\caption{Structure and Brillouin zone (BZ) of bulk and 2D Fe$_3$GeTe$_2$. 
(a) Side view of the hexagonal bulk structure with two Fe$_3$GeTe$_2$ monolayers per unit cell and 
(b) the corresponding hexagonal BZ. 
(c) Top view of one Fe$_3$GeTe$_2$ monolayer whose side view is indicated by 
the red dashed rectangle in (a). 
The $\Gamma$-K-M plane in (b) can be regarded as the corresponding 2D BZ.
Note that in (c), the FeII triangular lattice sits right underneath the Te lattice 
and thus cannot be seen here.
}
\label{fig:struc}
\end{figure}

\section{STRUCTURES AND METHODS}
Bulk Fe$_3$GeTe$_2$ forms a layered hexagonal structure with space group P6$_3$/mmc 
(No. 194) and point group $D_{6h}$, and its experimental lattice constants are $a=3.991$ {\AA} 
and $c=16.33$ {\AA}~\cite{Deiseroth2006}.
Each unit cell contains two weak interacting Fe$_3$GeTe$_2$ MLs [see Fig. \ref{fig:struc}(a)].
Each Fe$_3$GeTe$_2$ ML consists of five 2D triangular atomic layers where two Te lattices sandwich 
three ABA-stacked Fe lattices [see Fig. \ref{fig:struc}(a)] with 
the Fe and Ge triangular lattices in the central
layer forming a honeycomb lattice [see Fig. \ref{fig:struc}(c)]. 
The Fe atoms sit on two inequivalent sites, namely, the FeI site
with $C_{3v}$ site symmetry and the FeII site with $D_{3h}$ site symmetry [see Fig. \ref{fig:struc}(a)].

The $ab$ $initio$ electronic structure and structural optimization calculations are based on 
the density functional theory with the generalized gradient approximation (GGA)~\cite{Perdew1996}
to exchange-correlation interaction. To accurately describe the structural properties of layered Fe$_3$GeTe$_2$
structures, we have included the DFT-D2 vdW correction of Grimme \cite{Grimme2006} in the present calculations. 
The present calculations are performed using the accurate projector-augmented wave (PAW) method 
\cite{Blochl1994} implemented in the Vienna $ab$ $initio$ simulation package (VASP) \cite{Kresse1993,Kresse1996}. 
The few-layer Fe$_3$GeTe$_2$ structures are modelled utilizing the slab-superlattice approach
with the separations between the neighboring slabs being about 15 \AA.
A large plane wave cut-off energy of 400 eV is used. For the Brillouin zone integrations 
using the tetrahedron method \cite{Jepson1971}, $\Gamma$-centered $k$-meshes of 16 $\times$ 16 $\times$ 4 
and 16 $\times$ 16 $\times$ 1 are used for bulk and few-layer Fe$_3$GeTe$_2$, respectively. 
Since the experimental structural parameters for few-layer Fe$_3$GeTe$_2$ are unavailable, 
we have determined theoretically the lattice constants and atomic positions 
of both bulk and few-layer Fe$_3$GeTe$_2$.
The calculated lattice constants are listed in Table I, together with the experimental lattice constants
of bulk Fe$_3$GeTe$_2$. Note that the calculated lattice constants of  bulk Fe$_3$GeTe$_2$
agree well with the corresponding experimental values (within 0.5 \%) (see Table I), 
suggesting that the structural properties of few-layer Fe$_3$GeTe$_2$ should be well described by
the GGA functional~\cite{Perdew1996} plus the DFT-D2 vdW correction~\cite{Grimme2006}.

\begin{table}[htbp]
\label{table:1}
\caption{Theoretical lattice constants and crystallographic point group of bulk and few-layer Fe$_3$GeTe$_2$.
Experimental lattice constants
of bulk Fe$_3$GeTe$_2$~\cite{Deiseroth2006} are also listed in brackets for comparison.
Here $d$ is the effective thickness of few-layer Fe$_3$GeTe$_2$}
\begin{ruledtabular}
\begin{tabular}{c c c c c c c}
&Bulk & ML & BL & TL & FL & QL \\
\hline
Point Group & D$_{6h}$& D$_{3h}$& D$_{3d}$ & D$_{3h}$ & D$_{3d}$ & D$_{3h}$ \\
$a$ (\AA)& 3.995 (3.991$^a$) & 4.010 & 4.010 & 4.010 & 4.010 & 4.010 \\
$c$ (\AA)& 16.73 (16.33$^a$) & & & & & \\
$d$ (\AA)&     & 8.365 & 16.73 & 25.10 & 33.46 & 41.83   
\end{tabular}
\end{ruledtabular}
{$^{a}$Reference \cite{Deiseroth2006}  (X-ray diffraction experiment at 293K).} 
\end{table}

To determine the MCE, we first perform the total energy calculations for an in-plane 
magnetization (e.g., along the $x$-axis) and the out-of-plane magnetization (along the $z$-axis).
Then, the MCE is given by the total energy difference $\Delta E_c=E^{[100]}-E^{[001]}$. 
Denser $k$-point meshes of 24 $\times$ 24 $\times$ 6 and 32 $\times$ 32
$\times$ 1 are used for bulk and few-layer Fe$_3$GeTe$_2$, respectively.
Thus-obtained MCEs are converged within 3 \% with respect to the $k$-point meshes used. 

Similarly, the MDE is given by the difference in the magnetic dipole interaction energy ($E_d$)
between the magnetization along the $x$-axis and along the $z$-axis. For 
a ferromagnetic system, in atomic Rydberg units,~\cite{Guo1991A,Guo2007}
\begin{equation}
E_d=\sum_{qq'}\frac{2m_q m_{q'}}{c^2} M_{qq'}
\end{equation}
where the speed of light $c = 274.072$ and the so-called magnetic dipolar Madelung constant 
\begin{equation}
M_{qq'}=\sum_{\boldsymbol{R}}\frac{1}{\boldsymbol{|}\boldsymbol{R}+\boldsymbol{q}+\boldsymbol{q'}\boldsymbol{|}^3}
\left\{1-3\frac{[(\boldsymbol{R}+\boldsymbol{q}+\boldsymbol{q'})\cdot\hat{m}_q]^2}{\boldsymbol{|}\boldsymbol{R}+
\boldsymbol{q}+\boldsymbol{q'}\boldsymbol{|}^2}\right\}   
\end{equation}
where $\boldsymbol{R}$ are the lattice vectors, $\boldsymbol{q}$ are the atomic position vectors in the unit cell, and 
$m_q$ is the atomic magnetic moment (in units of $\mu_B$) on site $q$. 
In a 2D system, all $\boldsymbol{R}$ and $\boldsymbol{q}$ are in-plane,  
and hence the second term in Eq. (2) would vanish for the out-of-plane magnetization. 
Thus, the $E_d$ for an in-plane magnetization is always lower than that for
the out-of-plane magnetization. Consequently, the MDE always prefers an in-plane magnetization 
rather than the out-of-plane magnetization in 2D  magnetic systems~\cite{Guo1991A,Yimei2018,Vijay2019}. 
This is a purely geometrical effect and thus MDE is also called the magnetic 
shape anisotropy energy. 

For a solid with at least threefold rotational symmetry and a magnetization along rotational $z$-axis, 
the non-zero components for the optical conductivity are $\sigma_{xx}$, $\sigma_{zz}$, and $\sigma_{xy}$. 
We calculate these three independent elements using the linear-response Kubo formula
\cite{Wang1974,Oppeneer1992,Feng2015}. Thus, the absorptive parts of these elements are given by
\begin{equation}
\sigma^1_{aa}(\omega)=\frac{\pi e^2}{\hbar\omega m^2}\sum_{i,j}\int_{BZ}\frac{d\boldsymbol{k}}{(2\pi)^3}\left| 
p^a_{ij}\right|^2\delta(\epsilon_{\boldsymbol{k}j}-\epsilon_{\boldsymbol{k}i}-\hbar\omega),
\end{equation}
\begin{equation}
\sigma^2_{xy}(\omega)=\frac{\pi e^2}{\hbar\omega m^2}\sum_{i,j}\int_{BZ}\frac{d\boldsymbol{k}}{(2\pi)^3} Im
\left[p^x_{ij}p^y_{ij}\right]\delta(\epsilon_{\boldsymbol{k}j}-\epsilon_{\boldsymbol{k}i}-\hbar\omega),
\end{equation}
where $\hbar\omega$ is the photon energy and $\epsilon_{\boldsymbol{k}i}$ is the $i$th band energy at point $
\boldsymbol{k}$. Summations $i$ and $j$ are over the occupied and unoccupied bands, respectively. Dipole matrix 
elements $p^a_{ij}=\left\langle\boldsymbol{k}j\left|\hat{p}_a\right|\boldsymbol{k}i\right\rangle$, where $\hat{p}_a$ 
denotes the Cartesian component $a$ of the dipole operator, are obtained from the relativistic band structure within 
PAW formalism \cite{Adolph2001}. The integration is over the whole Brillouin zone using the linear tetrahedron 
method \cite{Temmerman1989}. The dispersive part of the optical conductivity elements can be obtained 
from its corresponding absorptive parts using the Kramers-Kronig relations,
\begin{equation}
\sigma^2_{aa}(\omega)=-\frac{2\omega}{\pi}P\int^\infty_0\frac{\sigma^1_{aa}(\omega')}{\omega'^2-\omega^2}d\omega',
\end{equation} 
\begin{equation}
\sigma^1_{xy}(\omega)=\frac{2}{\pi}P\int^\infty_0\frac{\omega'\sigma^2_{xy}(\omega')}{\omega'^2-\omega^2}d\omega',
\end{equation}
where $P$ denotes the principal value. 

For a bulk magnetic material, the complex polar Kerr rotation angle is given by 
\cite{Guo1995,Guo1994}
\begin{equation}
\theta_K+i\epsilon_K=\frac{-\sigma_{xy}}{\sigma_{xx}\sqrt{1+i(4\pi/\omega)\sigma_{xx}}}.
\end{equation} 
However, for a magnetic thin film on a nonmagnetic substrate, the complex polar Kerr rotation is given by 
\cite{Sivadas2016,Wu2019}
\begin{equation}
\theta_K+i\epsilon_K=\frac{-2(Z_0 d\sigma_{xy})}{(n_s+Z_0 d\sigma_{xx})^2-1}
\end{equation}
where $n_s$ is the refractive index of the substrate, $Z_0$ is the impedance of free space, 
and $d$ stands for the thickness of the magnetic layer.
Since few-layer Fe$_3$GeTe$_2$ 
were usually deposited on a SiO$_2$/Si substrate \cite{Deng2018,Fei2018}, the refractive index 
of bulk SiO$_2$ ($n_{s}=1.5$) is used here. 
Similarly, the complex Faraday rotation angle for a thin film can be written as \cite{Ravindran1999}
\begin{equation}
\theta_F+i\epsilon_F = \frac{\omega d}{2c}(n_+ - n_-),
\end{equation}
where $n_+$ and $n_-$ represent the refractive indices for left- and right-polarized lights, respectively. 
The refractive indices are related to the optical conductivity and the dielectric function via expression 
$n_\pm=\sqrt{\epsilon_\pm}=\sqrt{1+\frac{4\pi i} {\omega}\sigma_\pm}
=\sqrt{1+\frac{4\pi i}{\omega}(\sigma_{xx}\pm i\sigma_{xy})} 
\approx 1+\frac{4\pi i}{\omega}(\sigma_{xx}\pm \frac{i}{2}\sigma_{xy})$.
As a result, 
\begin{equation}
\theta_F+i\epsilon_F \approx -\frac{2\pi d}{c}\sigma_{xy}.
\end{equation}
Here the real parts of the optical 
conductivity $\sigma_\pm$ can be written as 
\begin{equation}
\sigma^1_\pm(\omega)=\frac{\pi e^2}{\hbar\omega m^2}\sum_{i,j}\int_{BZ}\frac{d\boldsymbol{k}}{(2\pi)^3}\left| \Pi^
\pm_{ij}\right|^2\delta(\epsilon_{\boldsymbol{k}j}-\epsilon_{\boldsymbol{k}i}-\hbar\omega),
\end{equation} 
where $\Pi^\pm_{ij}=\left\langle\boldsymbol{k}j\left| \frac{1}{\sqrt{2}}(\hat{p}_x\pm i\hat{p}_y)\right|
\boldsymbol{k}i\right\rangle$. Clearly, $\sigma_{xy}=\frac{1}{2i}(\sigma_+ - \sigma_-)$ and therefore $\sigma_{xy}$ 
would be nonzero only if $\sigma_+$ and $\sigma_-$ are different. In other words, magnetic circular dichroism is the 
fundamental cause of the nonzero $\sigma_{xy}$ and hence the MO effect. 

\begin{table*}
\label{table:2}
\caption{Calculated total spin magnetic moment ($m^s_t$), atomic (averaged) spin magnetic moment ($m^s_{FeI}$, $m^s_{FeII}$, 
$m^s_{Ge}$, $m^s_{Te}$), and orbital magnetic moment ( $m^o_{FeI}$, $m^o_{FeII}$, $m^o_{Ge}$, $m^o_{Te}$), 
as well as density of states at the Fermi level [$D(E_F)$],
magnetocrystalline anisotropy ($\Delta E_c$), magnetic dipolar anisotropy energy ($\Delta E_d$) and total 
magnetic anisotropy energy ($\Delta E_{ma}=\Delta E_c+\Delta E_d$) of bulk and few-layer Fe$_3$GeTe$_2$. Positive 
$\Delta E$ values mean that the out-of-plane magnetization is favored. The spin and orbital moments  
for two magnetization directions ({\bf m}$\parallel$[001] and {\bf m}$\parallel$[100]) are listed. 
The related experimental and theoretical values are listed for comparison.} 
\begin{ruledtabular}
\begin{tabular}{c c c c c c c c c c}  
System & & $D(E_F)$ & $m^s_t$& $m^s_{FeI}(m^o_{FeI})$& $m^s_{FeII}$ ($m^o_{FeII}$)& $m^s_{Ge}$ ($m^o_{Ge}$)& $m^s_{Te}$ 
($m^o_{Te}$)& $\Delta E_c$ ($\Delta E_d$)& $\Delta E_{ma}$\\
& &(1/eV/f.u.) & ($\mu_B/$f.u.)& ($\mu_B$/atom)& ($\mu_B$/atom)& ($\mu_B$/atom)& ($\mu_B$/atom)& (meV/f.u.)& (meV/f.u.) \\ 
\hline
bulk & {\bf m}$\parallel$[001] & 3.25 & 6.29 & 2.41 (0.075) & 1.53 (0.030) & -0.10 (0.001)& -0.04 (-0.016)& 3.37 (-0.086)& 3.28\\
     &           &                  & 4.95$^a$ & 2.18$^b$ (0.10$^c$)    & 1.54$^b$ (0.10$^c$)   & -& -& $\sim$3.4$^d$& -   \\
     & {\bf m}$\parallel$[100] & & 6.29& 2.41 (0.084)& 1.54 (0.050)& -0.10 (0.002)& -0.04 (-0.006)& -&  -\\
ML  & {\bf m}$\parallel$[001] &3.24& 6.27& 2.44 (0.076)& 1.47 (0.033)& -0.10 (0.001)& -0.04 (-0.018)& 3.00 (-0.32)& 2.68\\
         &                         &    & -   & 1.72$^e$    & 1.01$^e$ & -& -& 2.76$^e$(-0.11$^e$)& 2.0$^f$, 2.7$^g$  \\
         & {\bf m}$\parallel$[100] & & 6.28& 2.43 (0.085)& 1.49 (0.049)& -0.10 (0.003)& -0.04 (-0.007)& -& - \\
BL  & {\bf m}$\parallel$[001]  & 3.26 &6.31& 2.43 (0.076)& 1.54 (0.033)&  -0.10 (0.001)& -0.04 (-0.017)& 3.02 (-0.33)& 2.69\\
       & {\bf m}$\parallel$[100] &      & 6.31& 2.42 (0.085)& 1.54 (0.052)&  -0.10 (0.003)& -0.04 (-0.007)& -& -\\
TL & {\bf m}$\parallel$[001] & 3.34 & 6.31& 2.43 (0.076)& 1.54 (0.032)& -0.10 (0.001)& -0.04 (-0.017)& 3.20 (-0.33)& 2.87\\
        & {\bf m}$\parallel$[100] & & 6.32& 2.42 (0.085)& 1.55 (0.050)&  -0.10 (0.003)& -0.04 (-0.006)& -& -\\
FL & {\bf m}$\parallel$[001] & 3.34 & 6.31& 2.43 (0.076)& 1.54 (0.032)& -0.10 (0.001)& -0.04 (-0.017)& 3.22 (-0.32)& 2.90\\
        & {\bf m}$\parallel$[100] & & 6.32& 2.43 (0.084)& 1.55 (0.051)&  -0.10 (0.003)& -0.04 (-0.007)& -& -\\
QL & {\bf m}$\parallel$[001] & 3.15 & 6.32& 2.43 (0.076)& 1.54 (0.032)& -0.10 (0.001)& -0.04 (-0.017)& 3.30 (-0.32)& 2.98\\
        & {\bf m}$\parallel$[100] & & 6.32& 2.43 (0.084)& 1.55 (0.051)&  -0.10 (0.002)& -0.04 (-0.006)& -& -\\
\end{tabular}
\end{ruledtabular}
$^{a}$Reference\cite{Chen2013} (SQUID experiment); $^{b}$Reference \cite{May2016} (Neutron scattering experiment at 4K);
{$^{c}$Reference\cite{Zhu2016} (XMCD experiment: averaged orbital moment); $^{d}$Reference\cite{Park2020}(GGA); 
$^{e}$Reference \cite{Zhuang2016}(LDA);$^{f}$Reference \cite{Deng2018}(RMCD experiment);$^{g}$Reference \cite{Wang2020}(LDA)} 
\end{table*}

\section{RESULTS AND DISCUSSION}
\subsection{Magnetic moments and magnetic anisotropy energy}
Since both bulk and few-layer Fe$_3$GeTe$_2$ have been experimentally found to be 
ferromagnetic~\cite{Fei2018,Deng2018,Deiseroth2006,Chen2013}, 
we consider only the ferromagnetic configuration in this paper.
The calculated spin and orbital magnetic moments of the considered Fe$_3$GeTe$_2$ structures 
are listed in Table II, together with related previous experimental and theoretical results. 
First of all, Table II shows that calculated magnetic moments in few-layer Fe$_3$GeTe$_2$ 
hardly depend on their thickness (i.e., the number of MLs) and also they are very close
to that in bulk Fe$_3$GeTe$_2$. This can be expected from the fact that the interlayer interaction
is weak in the considered Fe$_3$GeTe$_2$ structures. 
As mentioned above, in all the considered systems there are two inequivalent Fe sites (FeI and FeII) 
with different site symmetries (see Fig. \ref{fig:struc}). Interestingly, Table II shows 
that in all the considered systems, FeI and FeII have rather different spin magnetic moments,
and the difference is nearly as large as 0.9 $\mu_B$. 
For example, the spin magnetic moments of FeI and FeII in bulk Fe$_3$GeTe$_2$ are 
2.41 and 1.53 $\mu_B$, respectively. In the ionic picture, the valence state of Fe$_3$GeTe$_2$
could be written as (FeII$^{2+}$)(FeI$^{3+}$)$_2$Ge$^{4-}$(Te$^{2-}$)$_2$~\cite{Deng2018} and thus we would expect
FeI ($d^5$) and FeII ($d^6$) to have the spin moments of 5.0 and 4.0 $\mu_B$, respectively. The fact that
calculated spin moments are significantly smaller than these values, supports the notion that
these Fe$_3$GeTe$_2$ structures are itinerant ferromagnets~\cite{Fei2018,Deng2018,Zhuang2016}.

We note that these theoretical spin magnetic moments are in good agreement with the neutron 
scattering data~\cite{May2016} (Table II). 
This indicates the validity of the GGA functional used here for describing
the magnetic properties of the considered Fe$_3$GeTe$_2$ structures.
We can expect that the total magnetic moment
in these Fe$_3$GeTe$_2$ structures comes mostly from the Fe atoms, and this would result
in a total magnetic moment of 6.35 $\mu_B$/f.u. The small difference between this value
and the calculated total spin moment is due to the small spin moments of Te and Ge
which are antiparallel to that of Fe (see Table II). 
We also notice that the calculated total magnetic moment is significantly larger than
that from the SQUID magnetization measurement (4.95 $\mu_B$/f.u.)~\cite{Chen2013}.
This may indicate that the samples used~\cite{Chen2013} were nonstoichiometric or
contained defects and Fe vacancies.
Due to the well-known crystal-field quenching, the calculated orbital magnetic moments of the Fe atoms
(e.g., 0.076 $\mu_B$/Fe for FeI and 0.033 $\mu_B$/Fe for II in bulk  Fe$_3$GeTe$_2$) 
are much smaller than the spin magnetic moments. The averaged calculated orbital moment of the Fe atoms
is $\sim$0.06 $\mu_B$/Fe, being smaller but in the same order of magnitude as 
that (0.10 $\mu_B$/Fe) from the XMCD measurement~\cite{Zhu2016}. It is known that the GGA and LDA functionals
would give rise to too small orbital moments by up to 40 \%  (see, e.g., ~\cite{Guo1997,Verchenko2015} and references therein).
Nonetheless, these discrepancies between the experiment and theory can be largely removed
by including the so-called orbital polarization correction in the electronic structure 
calculations (see, e.g., ~\cite{Guo1997,Verchenko2015} and references therein). 
Interestingly, Te orbital magnetic moments are comparable to Te spin magnetic moments 
and furthermore depend strongly on the magnetization direction (Table II). 
This is because the SOC in Te atoms is much stronger than in Fe and Ge atoms. 

We present the calculated magnetic anisotropy energies ($\Delta E$) in Table II. By definition ($\Delta E=E^{[100]}-E^{[001]}$), 
a positive $\Delta E$ value indicates an out-of-plane magnetization easy axis [i.e., the perpendicular magnetic anisotropy (PMA)]. 
Strikingly, Table II shows that all the considered Fe$_3$GeTe$_2$s structures have
a very large PMA, being $\sim$3.0 meV/f.u. or $\sim$1.0 meV/Fe. These $\Delta E_{ma}$ values not only
are several times larger than that of 2D ferromagnetic semiconductors  Cr$_2$Ge$_2$Te$_6$ 
($\sim$0.1 meV/f.u.) \cite{Yimei2018} and CrI$_3$ ($\sim$0.5 meV/f.u.)~\cite{Vijay2019} but also
are comparable to that of heavy metal magnetic alloys such as FePt (2.75 meV/Fe)~\cite{Oppeneer1998},
which is known to have the largest MAE among the transition metal alloys.
This large PMA would strongly suppress the magnetic fluctuations in these 2D materials
and thus result in a higher ferromagnetic ordering temperature ($T_c$) than 2D ferromagnetic 
semiconductors CrI$_3$~\cite{Huang2017} and Cr$_2$Ge$_2$Te$_6$~\cite{Gong2017}.
We note that the calculated MAE (2.9 meV/f.u.) of ML Fe$_3$GeTe$_2$ is  
in good agreement with the RMCD experimental data (2.0 meV/f.u.)~\cite{Deng2018}.
The largeness of the PMA plus its electric tunability~\cite{Deng2018} suggests 
that 2D ferromagnetic metallic Fe$_3$GeTe$_2$ 
would have promising applications in high-density magnetic data storage and other spintronic devices.

Table II indicates that the MAE per f.u. increases slightly as one moves from ML to BL and then TL.
However, the MAE per f.u. remains unchanged as the film thickness is further increased, e.g.,
to that of FL and QL (Table II). As mentioned before, the MAE consists of two competing contributions,
namely, the MCE, which prefers PMA, and MDE, which always favors an in-plane magnetization.
The MDE in 2D Fe$_3$GeTe$_2$ is in the same magnitude as that in 2D CrI$_3$ and Cr$_2$Ge$_2$Te$_6$.
However, the magnitude of the MCE in 2D Fe$_3$GeTe$_2$ is nearly ten times larger than that of the MDE.
This results in a gigantic MAE in 2D Fe$_3$GeTe$_2$. Note that these values are nearly three orders 
of magnitude larger than that of ferromagnetic Fe and Ni (~4 $\mu$eV/f.u.) \cite{Guo1991B}. 
Bulk Fe$_3$GeTe$_2$ has an even larger MAE of 3.28 meV/f.u.,
simply because its MDE is much smaller than that in 2D Fe$_3$GeTe$_2$ (Table II).
We notice that only one previous study, based on the local density approximation (LDA), 
on the MDE has been reported.~\cite{Zhuang2016} Nevertheless, the calculated MDE for ML Fe$_3$GeTe$_2$
in Ref. ~\cite{Zhuang2016} is nearly three times smaller than the present calculations (Table II).
This is mainly because the LDA spin magnetic moments reported in ~\cite{Zhuang2016} are
significantly smaller than the present GGA calculations (see Table II). 

Table II also shows that while the spin magnetic moments are generally independent of
the magnetization orientation, the orbital magnetic moments change significantly 
as the magnetization is rotated, e.g., from [001] to [100]. In particular, the Te orbital moment
in ML Fe$_3$GeTe$_2$ gets reduced by nearly 60 \% when the magnetization is rotated from [001] to [100],
while that of FeI increases by about 12 \% (Table II). 
This may suggest some correlation between the MCE and the anisotropy in the orbital magnetization.
Indeed, a previous perturbative theory analysis showed that in elemental transition metal MLs,
the easy axis is along the direction in which the orbital moment is largest~\cite{Bruno1989}.
This was supported by the {\it ab initio} calculations for Fe monolayers imbedded in noble metals~\cite{Guo1991A}.
Interestingly, the orbital magnetic moments on heavy element Te atoms (which have the largest SOC)
in all the considered Fe$_3$GeTe$_2$ structures with the perpendicular magnetization 
are much larger than that for an in-plane magnetization,
thus indicating a possible connection with the strong PMA in these systems.
Nevertheless, Table II shows that the Fe orbital moments in these Fe$_3$GeTe$_2$ structures are
larger for an in-plane magnetization than for the perpendicular magnetization,
indicating that the perturbative theory analysis for transition metal monolayers~\cite{Bruno1989}
may not be wholely applicable to the present ternary compounds.

\begin{figure*}[tbph] \centering
\includegraphics[width=14cm]{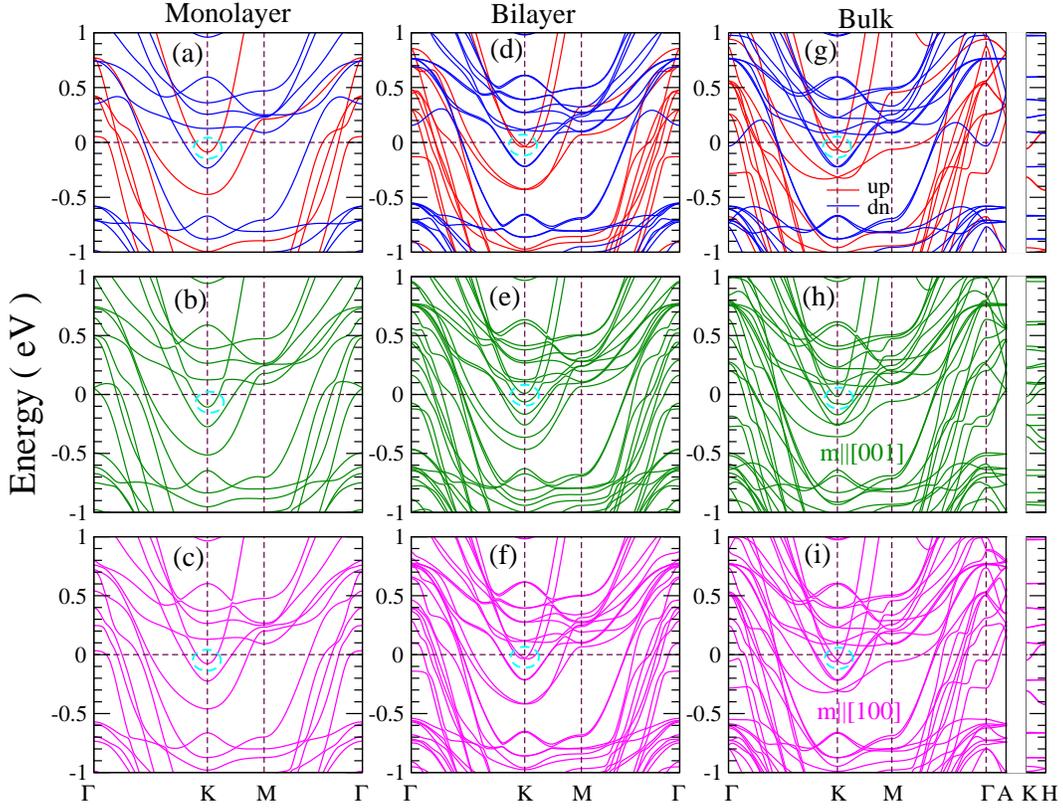}
\caption{Band structures of ML (left column), BL (middle column) and bulk (right column) Fe$_3$GeTe$_2$.
Spin-polarized scalar-relativistic band structures (upper row), relativistic band structures with the magnetization
along $z$-axis (middle row) and with an in-plane magnetization along $x$-axis. In (d-i), the dotted ellipse circles
the topological node [gapless: (d,f,g,i) and gapped: (e,h)]. The Fermi level is at 0 eV.}
\label{fig:mlbband}
\end{figure*}

\subsection{Electronic structure}

To understand the calculated magnetic and optical properties of the Fe$_3$GeTe$_2$ materials, 
we present the calculated electronic band structures.  The band structures of ML, BL and bulk Fe$_3$GeTe$_2$ 
are displayed in Fig. \ref{fig:mlbband}, and that of TL, FL and QL Fe$_3$GeTe$_2$ in Fig. S1 in
the supplementary material (SM)~\cite{SM}. Since the band structures of bulk and ML Fe$_3$GeTe$_2$ 
have already been reported in several previous papers~\cite{Deng2018,Zhuang2016}
and are also rather similar to that of TL, FL and QL Fe$_3$GeTe$_2$ due to the weak interlayer
interaction, here we only summarize the salient 
features of the calculated band structures.

First of all, all the considered Fe$_3$GeTe$_2$ structures are metallic with multiple Fermi surface pockets
(Fig. \ref{fig:mlbband} and Fig. S1 in the SM~\cite{SM}) and hence a large density of states (DOS) 
at the Fermi level ($E_F$) (Table II). In all the considered systems, there are many hole Fermi surface 
pockets centered at the $\Gamma$ point and many electron Fermi surface pockets centered 
at the K point in the BZ (Figs. \ref{fig:mlbband} and Fig. S1 in the SM). 
Secondly, because bulk Fe$_3$GeTe$_2$ contains two MLs per unit cell and also
the interlayer interaction is weak, the band structure of BL Fe$_3$GeTe$_2$ [Figs. 2(d), 2(e) and 2(f)]
are almost identical to that of bulk Fe$_3$GeTe$_2$ [Figs. 2(g), 2(h) and 2(i)].
The band structure of ML Fe$_3$GeTe$_2$ is very similar to that of BL Fe$_3$GeTe$_2$ (Fig. 2)
except that the number of the bands is only half of that for the BL.
Similarly, the band structures of TL, FL and QL Fe$_3$GeTe$_2$ 
are overall nearly the same as that of the BL except with the increased number of bands (Fig. S1 in the SM).

\begin{figure}[t] \centering
\includegraphics[width=8.6cm]{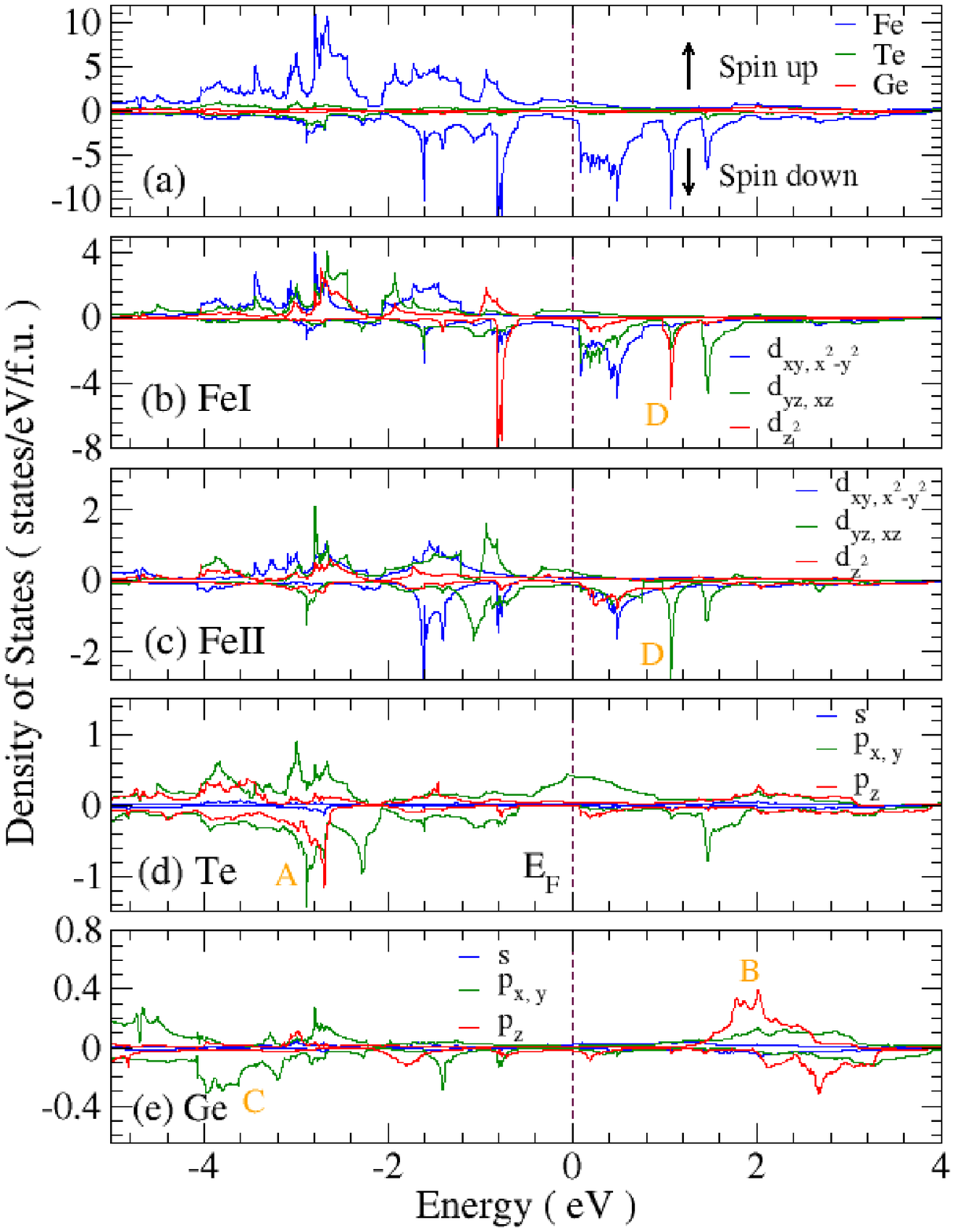}
\caption{Scalar-relativistic site-, orbital-, and spin-projected DOS of bulk Fe$_3$GeTe$_2$. The Fermi 
level is at 0 eV.} 
\label{fig:bdos}
\end{figure}

\begin{figure}[t] \centering
\includegraphics[width=8.6cm]{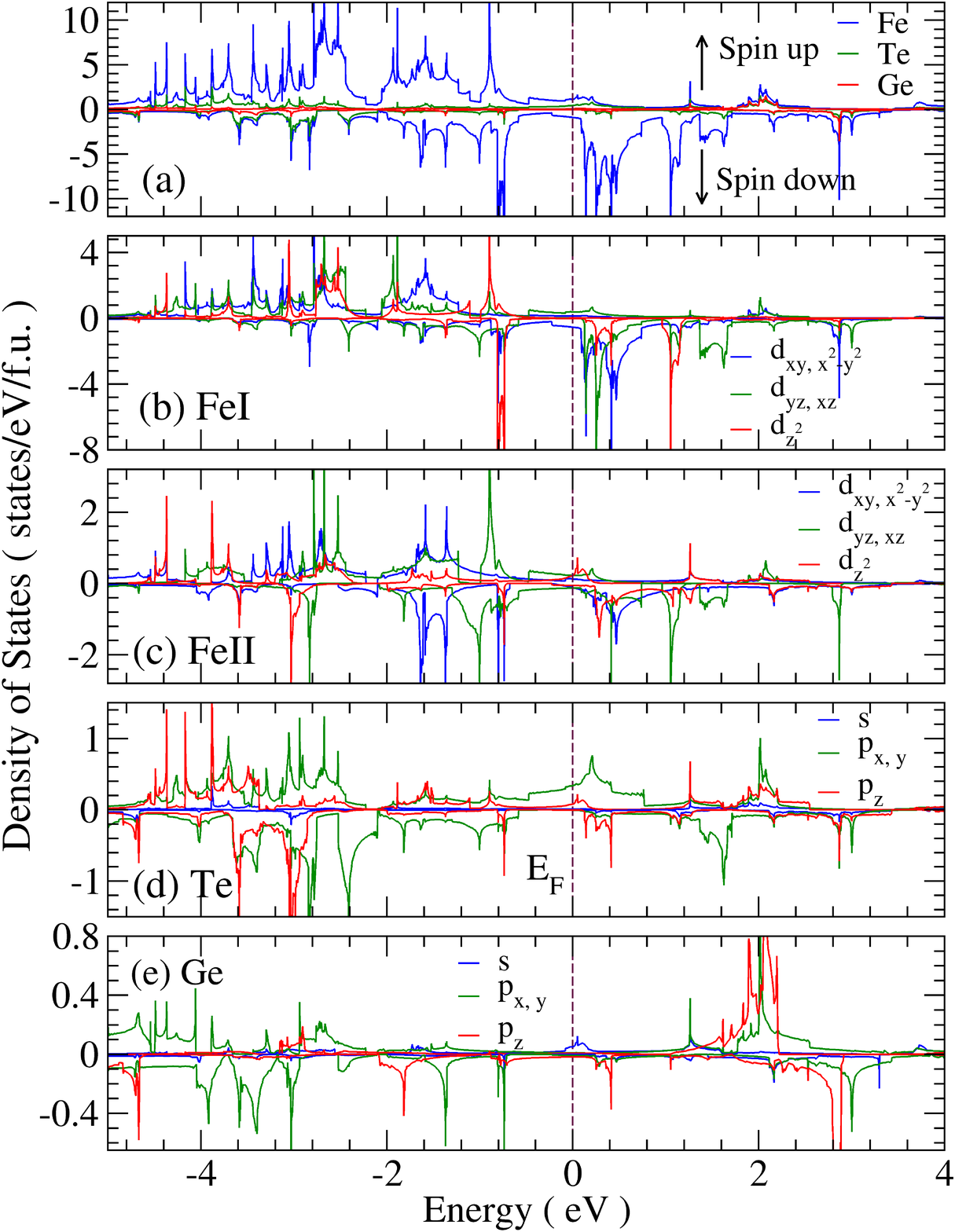}
\caption{Scalar-relativistic site-, orbital-, and spin-projected DOS of monolayer Fe$_3$GeTe$_2$. The Fermi 
level is at 0 eV.} 
\label{fig:mldos}
\end{figure}

Next, we present the total as well as site-, orbital-, and spin-projected DOS spectra 
of bulk and 2D Fe$_3$GeTe$_2$ in Figs. 3 and 4 as well as Figs. S2 and S3 in the SM~\cite{SM}. 
Figure \ref{fig:bdos} shows that for bulk Fe$_3$GeTe$_2$,
the lower valence bands ranging from -5.0 to -2.2 eV result mainly from the hybridization 
among Fe $d$, Ge $p$, and Te $p$-orbitals. The upper valence bands and lower conduction bands 
(from -2.2 to 1.0 eV) are dominated by Fe $d$-orbitals together with minor contributions of 
Te $p$-orbitals around the Fermi level. 
Furthermore, Figs. \ref{fig:bdos}(b) and \ref{fig:bdos}(c) show that spin-splitting of Fe $d$ bands is large, being 
more than 1 eV and thus indicating strong intra-atomic exchange interaction in bulk Fe$_3$GeTe$_2$. 
Interestingly, the local DOS spectra for the FeI and FeII sites are rather different,
being consistent with the rather different magnetic moments on these sites (Table II).
These differences are caused by their different local environments (coordination number and site symmetry).
For example, the main spin-down FeI $d_{z^2}$ DOS peak is very sharp and is located at -0.8 eV [Fig. \ref{fig:bdos}(b)]
while the main spin-down FeII $d_{z^2}$ DOS peak is rather broad and is centered at 0.4 eV [Fig. \ref{fig:bdos}(b)]. 
This is because there is no ligand atom lying above or below FeI [Figs. \ref{fig:struc}(a) and \ref{fig:struc}(c)]
and thus FeI $d_{z^2}$ orbitals form a rather localized narrow band.
In contrast, there is one Te atom sitting right above (or right below) the FeII atom [Figs. \ref{fig:struc}(a)].
Thus, FeII $d_{z^2}$- and Te $p_z$-orbitals hybridize and pushes FeII $d_{z^2}$-orbital dominated antibonding
band above the Fermi level [Fig. \ref{fig:bdos}(c)].

\begin{table*}
\label{table:3}
\caption{The SOC matrix elements $\langle m\sigma\left | \vec{L}\cdot\vec{S}\right |m'\sigma'\rangle$ in the $p$-orbital basis.
Here $\eta$ is the polar angle and $\phi$ is the azimuthal angle. For the perpendicular magnetization,
$\eta = \phi = 0$ while for the in-plane magnetization along the $x$-axis, $\eta = \pi/2$ and $\phi = 0$. 
}
\begin{ruledtabular}
\begin{tabular}{c c c c}
& $| p_x;\uparrow \rangle$ & $| p_y;\uparrow \rangle$ & $| p_z;\uparrow \rangle$ \\
\hline
\\[-0.8em]
$| p_x;\uparrow \rangle$  & 0& $-\frac{i}{2}\cos\eta$& $\frac{i}{2}\sin\eta\sin\phi$  \\
\\[-0.8em]
$| p_y;\uparrow \rangle$  & $\frac{i}{2}\cos\eta$& 0& $-\frac{i}{2}\sin\eta\cos\phi$  \\
\\[-0.8em]
$| p_z;\uparrow \rangle$  & $-\frac{i}{2}\sin\eta\sin\phi$& $\frac{i}{2}\sin\eta\cos\phi$& 0 \\
\\[-0.8em]
$| p_x;\downarrow \rangle$  & 0& $\frac{i}{2}\sin\eta$& $-\frac{1}{2}(\cos\phi-i\cos\eta\sin\phi)$ \\
\\[-0.8em]
$| p_y;\downarrow \rangle$  & $-\frac{i}{2}\sin\eta$& 0& $-\frac{1}{2}(\sin\phi+i\cos\eta\cos\phi)$ \\
\\[-0.8em]
$| p_z;\downarrow \rangle$  & $\frac{1}{2}(\cos\phi-i\cos\eta\sin\phi)$& $\frac{1}{2}(\sin\phi+i\cos\eta\cos\phi)$& 0  \\
\end{tabular}
\end{ruledtabular}
\end{table*}

Let us now discuss the possible origin of the large MCE in the Fe$_3$GeTe$_2$ systems in terms
of the Fe $d$-orbital decomposed DOS spectra in, e.g., Fig. \ref{fig:bdos}. 
Accoring to perturbation theory analysis, the occupied and empty $d$ states near the Fermi level 
are coupled by the SOC and thus make most important contributions to the MCE~\cite{Wang1993}.
Moreover, for the same spin channel, the SOC matrix elements $\langle d_{xy} \left | H_{SO}\right | d_{x^2-y^2}\rangle$ 
and $\langle d_{yz}\left | H_{SO}\right | d_{xz}\rangle$ prefer 
the out-of-plane anisotropy, while $\langle d_{yz} \left | H_{SO}\right | d_{xy}\rangle$, 
$\langle d_{yz}\left | H_{SO}\right | d_{z^2}\rangle$, and $\langle d_{yz}\left | H_{SO} \right | d_{x^2-y^2}\rangle$ 
favour an in-plane anisotropy \cite{Takayama1976}.
The ratios of these matrix elements are $\langle d_{xy}\left | H_{SO}\right | d_{x^2-y^2}\rangle^2$ 
: $\langle d_{yz}\left | H_{SO}\right | d_{xz}\rangle^2$ : $\langle d_{yz}\left | H_{SO}\right | d_{xy}\rangle^2$ 
: $\langle d_{yz}\left | H_{SO}\right | d_{z^2}\rangle^2$ : 
$\langle d_{yz}\left | H_{SO}\right | d_{x^2-y^2}\rangle^2$ = 4 : 1 : 1 : 3 : 1.~\cite{Takayama1976}
Figure \ref{fig:bdos}(b) shows that FeI $d_{yz,xz}$ and $d_{xy,x^2-y^2}$ orbitals dominate FeI $d$-orbital decomposed
DOS spectra in spin-up and spin-down channels, respectively. Consequently, the SOC matrix 
elements $\langle d_{xy}\left | H_{SO}\right | d_{x^2-y^2}\rangle$ and $\langle d_{yz}\left | H_{SO}\right | d_{xz}\rangle$ 
would make dominating contributions to the MCE, thereby giving rise to the large PMA in bulk Fe$_3$GeTe$_2$.
For the FeII sites, the situation is more complicated [Fig. \ref{fig:bdos}(c)]. 
In addition to the prominent $d_{yz,xz}$ orbitals, there are the pronounced $d_{z^2}$ orbital
in both spin channels and also the prominent spin-down $d_{xy,x^2-y^2}$ orbitals.
Consequently, there are competing contributions to the MCE from the SOC matrix elements 
of $\langle d_{xy}\left | H_{SO}\right | d_{x^2-y^2}\rangle$ and $\langle d_{yz}\left | H_{SO}\right | d_{xz}\rangle$, 
which prefers the PMA, and $\langle d_{yz}\left | H_{SO}\right | d_{z^2}\rangle$, which favors an in-plane magnetization.
However, since the ratios of these matrix elements are $\langle d_{xy}\left | H_{SO}\right | d_{x^2-y^2}\rangle^2$
: $\langle d_{yz}\left | H_{SO}\right | d_{xz}\rangle^2$ : $\langle d_{yz}\left | H_{SO}\right | d_{xy}\rangle^2$ : 
$\langle d_{yz}\left | H_{SO}\right | d_{z^2}\rangle^2$ 
= 4 : 1 : 1 : 3, we believe that FeII atoms would make positive contributions to the PMA although
they may be much smaller than FeI atoms.

Figures \ref{fig:bdos} and \ref{fig:mldos} show that there are pronounced occupied and 
unoccupied $p$-orbital DOSs from heavy Te atoms in the vicinity of $E_F$. 
To examine the possible contribution of these $p$-orbital states to the MCE, following ~\cite{Takayama1976} 
we derive the SOC matrix elements in the basis of $p$ orbitals, which are listed in Table III.  
Among the nonzero SOC elements, we find that $\langle p_x;\uparrow\left | H_{SO}\right | p_y;\uparrow\rangle$ 
prefers the out-of-plane magnetization but $\langle p_y(p_x);\uparrow \left |H_{SO}\right | p_z;\uparrow\rangle$ 
favors an in-plane magnetization. Furthermore, $\langle p_x;\downarrow \left |H_{SO}\right | p_y;\uparrow\rangle$
prefers an in-plane magnetization while $\langle p_x(p_y);\downarrow \left |H_{SO}\right | p_z;\uparrow\rangle$ has no preference.
Unlike that for $d$ orbitals~\cite{Takayama1976}, the expectation values of all nonzero elements 
$\langle p;\uparrow\downarrow \left |H_{SO}\right | p;\uparrow\downarrow\rangle^2$ are identical.
Interestingly, Fig. \ref{fig:bdos} and \ref{fig:mldos} show that the Te $p_x$ and $p_y$ make
a prominent contribution to the spin-up DOSs near the Fermi level. 
Since $\langle p_x;\uparrow \left |H_{SO}\right | p_y;\uparrow\rangle$ prefers the perpendicular anisotropy
and also Te atoms have a much stronger SOC than Fe atoms, the large MCE found in
bulk and 2D Fe$_3$GeTe$_2$ may result mainly from the presence of
of pronounced  Te $p_{x,y}$-orbital DOSs near the Fermi level. 

The site-, orbital- and spin-projected DOS spectra of ML Fe$_3$GeTe$_2$ are shown in Fig. \ref{fig:mldos}. All the features of
the DOS spectra are similar to that of bulk Fe$_3$GeTe$_2$ (Fig. \ref{fig:bdos}). One pronounced difference is the
appearance of FeII $d_{z^2}$ orbital dominated peaks near -3.0 eV in the lower valence bands. 
This can be attributed to the lack of the interlayer coupling because there is no Fe$_3$GeTe$_2$ ML above and below it. 
The absence of the hybridization among FeII $d_{z^2}$ orbitals from neighboring MLs (via Te $p_{z}$ orbitals) makes
FeII $d_{z^2}$ orbitals localized and the corresponding band narrow.
On the contrary, the peaks of FeI $d_{z^2}$ orbitals in bulk and ML Fe$_3$GeTe$_2$ are nearly the same 
because no ligand atom sits along the $c$ direction. 
The site-, orbital- and spin-projected DOSs of BL and TL Fe$_3$GeTe$_2$ are displayed in Figs. S2 and S3 
in the SM~\cite{SM}, where the spectral features fall in between that of ML and bulk Fe$_3$GeTe$_2$.

Finally, we notice a topological nodal point just below $E_F$ at the K point
in the scalar-relativistic band structure of bulk Fe$_3$GeTe$_2$ [Fig. 2(g)]. This nodal point
extends along the K-H line, and thus forms a topological nodal line.~\cite{Kim2018}
Interestingly, for the perpendicular magnetization, this nodal line becomes split by a 
large band gap of $\sim$60 meV when the SOC is switched-on [Fig. 2(h)]. 
This results in the upper band being pushed upwards nearly above
$E_F$ at the K point and the lower band moving downwards [Fig. 2(h)]. 
Furthermore, as reported before~\cite{Kim2018}, these SOC-split
bands have very large Berry curvatures with opposite signs near the nodal points, thus leading
to the large observed anomalous Hall effect in bulk Fe$_3$GeTe$_2$.~\cite{Kim2018}
Here we want to emphasize that this gap-opening at the nodal point also lowers the total band energy.
On the other hand, these nodal points remain gapless when the magnetization is in-plane
and hence the total band energy would remain much unchanged. Clearly, this would give rise to
a significant contribution to the large MCE in bulk Fe$_3$GeTe$_2$ (see Table II).
We notice that all 2D Fe$_3$GeTe$_2$ structures except ML Fe$_3$GeTe$_2$ have such a nodal point
at the K point [see Fig. 2 and Fig. S1 in the SM]. Therefore, we may conclude that to some extent,
the large MCE found in bulk and 2D Fe$_3$GeTe$_2$ also originate from the gap-opening of
the topological nodal point at the K point in these structures.
We also notice that in ML Fe$_3$GeTe$_2$, although there is no such a nodal point,
the top valence band at the K point is lowered by $\sim$19 meV when the magnetization is
out-of-plane [Fig. 2(b)] but goes up by 9 meV when the magnetization becomes
in-plane. Clearly, such band movements due to the presence of the SOC and the change 
of magnetization direction would also contribute significantly to the large MCE 
in ML Fe$_3$GeTe$_2$.

\subsection{Optical and magneto-optical conductivity}

\begin{figure}[tbph] \centering
\includegraphics[width=8.6cm]{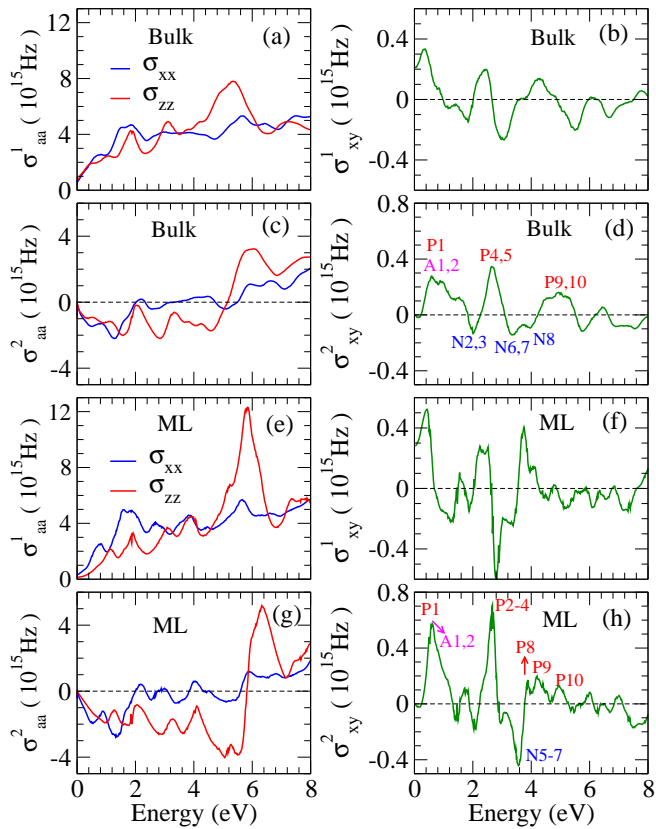}
\caption{Real (a)[(e)] diagonal and (b)[(f)] off-diagonal, imaginary (c)[(g)] diagonal components and (d)[(h)] off-
diagonal components of the optical conductivity tensor of bulk [ML] Fe$_3$GeTe$_2$ in the ferromagnetic state 
with the out-of-plane magnetization. All the spectra have been convoluted with a Lorentzian of 0.1 eV to simulate the finite 
electron lifetime effects.} 
\label{fig:bmloptics}
\end{figure}

We present the calculated optical and MO conductivity tensors for bulk and ML Fe$_3$GeTe$_2$ in Fig. \ref{fig:bmloptics}, 
and also for BL and TL as well as for FL and QL Fe$_3$GeTe$_2$ in Figs. S4 and S4 in the SM~\cite{SM}, respectively. 
First of all, we can see that all spectra are rather similar.
This similarity is due to the weak interlayer interaction in these materials. 
Such behaviour has also been reported in other 2D magnetic materials such as few-layer CrI$_3$~\cite{Vijay2019} 
and Cr$_2$Ge$_2$Te$_6$~\cite{Yimei2018}. Therefore, below we will analyze only the main features
in the optical and MO conductivity spectra of ML and bulk Fe$_3$GeTe$_2$ 
in detail as representatives of  these materials. 

Figures \ref{fig:bmloptics}(a) and \ref{fig:bmloptics}(e) show that the diagonal elements of the optical 
conductivity $\sigma_{xx}$ and $\sigma_{zz}$ differ significantly. Since the $\sigma_{xx}$ ($\sigma_{zz}$) is
the linear optical response of the materials to an in-plane electric field polarization $E\perp c$ 
(out-of-plane electric field polarization $E||c$), such difference indicates a large optical anisotropy. 
This can be expected for 2D or quasi-2D materials~\cite{Yimei2018,Vijay2019}. Specifically, 
the absorptive part of the diagonal element ($\sigma^1_{xx}$) for bulk Fe$_3$GeTe$_2$ (Fig. \ref{fig:bmloptics}(a)) 
is larger than $\sigma^1_{zz}$ in the low-energy region of 0.0-3.0 eV, but becomes smaller than  $\sigma^1_{zz}$in 
the high-energy region of 3.0-6.0 eV. Compared to that of bulk, ML Fe$_3$GeTe$_2$ has a similar difference 
between $\sigma^1_{xx}$ and $\sigma^1_{zz}$. Nevertheless, as expected, such difference gets slightly enhanced, 
because ML Fe$_3$GeTe$_2$ is a truly 2D material. 
We notice that this optical anisotropy can be further understood in terms of the calculated orbital-projected 
DOS spectra. Note that the $p_{x,y}$ and $d_{xy,x^2-y^2}$ ($p_z$ and $d_{z^2}$) states can only be excited 
by $E\perp c$ ($E || c$) polarized light while the $d_{xz,yz}$ states can be excited by both. 
Figures \ref{fig:bdos} and \ref{fig:mldos} indicate that the upper valence bands stem mainly
from Fe $d$ as well as Ge and Te $p$ orbitals. In the energy region between -3.0 and 0.0 eV, 
the overall weight of Fe $d_{xy,x^2-y^2}$ orbitals (excited by $E \perp c$) is slightly larger 
than that of Fe $d_{z^2}$ orbital (excited by $E || c$), thus leading to the slightly higher peaks 
of $\sigma^1_{xx}$ in the photon energy region below 3.0 eV 
[Figs. \ref{fig:bmloptics}(a) and \ref{fig:bmloptics}(e)].  
Further, broad peaks of Te $p_{x,y}$ DOS in the energy range from -1.0 to 1.0 eV 
[Figs. \ref{fig:bdos} and \ref{fig:mldos}(d)] also contribute to the higher peaks of $\sigma^1_{xx}$ 
below 3.0 eV. On the other hand, significant FeII $d_{z^2}$ and Te $p_z$ states appear 
at lower valance bands from -5.0 to -3.0 eV. This would explain the increase of $\sigma^1_{zz}$ ($E||c$) above 4.4 eV. 
Interestingly, among all considered structures, ML Fe$_3$GeTe$_2$ has a particularly sharp Te $p_z$ peak 
[see Figs.  \ref{fig:mldos}(c) and \ref{fig:mldos}(d)] due to the lack of interlayer coupling, 
as described above. This results in the largest optical anisotropy in 
ML Fe$_3$GeTe$_2$.        

The calculated real ($\sigma^1_{xy}$) and imaginary ($\sigma^2_{xy}$) parts of the off-diagonal optical conductivity 
elements are displayed in Figs. \ref{fig:bmloptics}(b) and \ref{fig:bmloptics}(b) (d) for bulk 
Fe$_3$GeTe$_2$ and in Figs. \ref{fig:bmloptics}(f) and \ref{fig:bmloptics}(h) for ML Fe$_3$GeTe$_2$. 
First of all, we note that in the DC-limit ($\omega\rightarrow 0$), $\sigma^1_{xy}(0)$ is actually 
the anomalous Hall conductivity (AHC). The AHCs calcuated this way are 
233 S/cm, 312 S/cm, 241 S/cm and 287 S/cm for bulk, ML, BL, and TL Fe$_3$GeTe$_2$, respectively. 
The AHC value for bulk Fe$_3$GeTe$_2$ agrees quite well with 
the experimental AHC value of $\sim$360 S/cm \cite{Xu2019} and also previous theoretical one 
of $\sim$287 S/cm \cite{Wang2019}. Secondly, as expected, the overall features in the $\sigma_{xy}$
spectra for bulk and ML Fe$_3$GeTe$_2$ are similar. 
For example, for both structures, the $\sigma_{xy}(\omega)$ spectra oscillates significantly 
with several high peaks. Prominent peaks occur mostly between 0.0 and 4.4 eV and 
beyond 4.4 eV, the magnitude of $\sigma_{xy}(\omega)$ get significantly reduced, 
indicating weak magnetic circular dichroism. Specifically, $\sigma^1_{xy}$ for ML Fe$_3$GeTe$_2$ 
has pronounced positive peaks at 0.4 eV and 3.8 eV as well 
as well as a negative peak at 2.8 eV [Fig.  \ref{fig:bmloptics}(f)]. 
On the other hand, $\sigma^2_{xy}$ for ML Fe$_3$GeTe$_2$ has large positive 
peaks at 0.6 eV and 2.6 eV as well as a large negative peak at 3.6 eV [Fig.  \ref{fig:bmloptics}(h)]. 
In comparison, the $\sigma_{xy}(\omega)$ spectra of bulk Fe$_3$GeTe$_2$ have peak 
positions and shapes being rather similar to that of ML but with significantly reduced magnitudes
[Figs.  \ref{fig:bmloptics}(b) and \ref{fig:bmloptics}(d)]. 

\begin{figure}[t] \centering
\includegraphics[width=8.6cm]{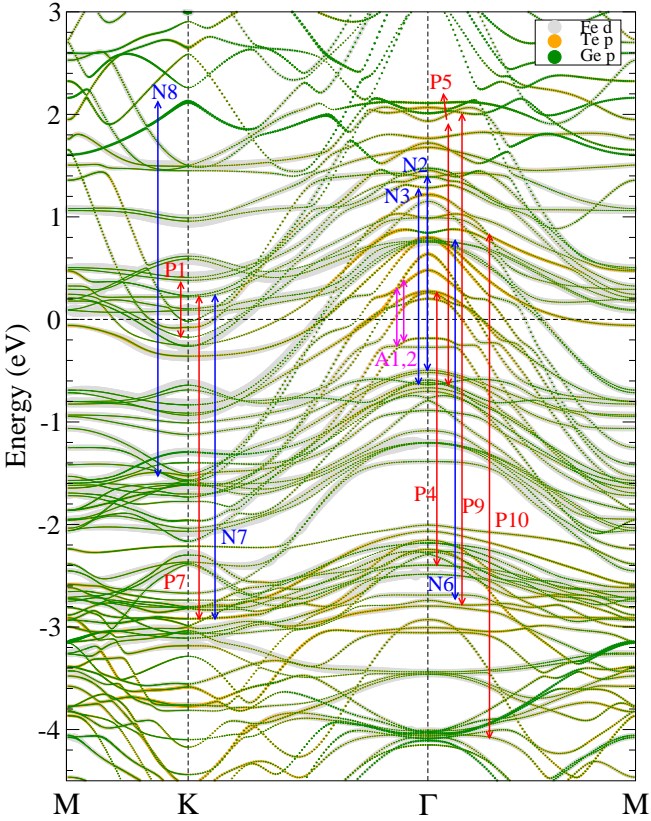}
\caption{Relativistic site- and orbital-projected band structures of bulk ferromagnetic Fe$_3$GeTe$_2$. The Fermi 
level is at 0 eV. The main interband transitions at the $\Gamma$ and K point, as well as the corresponding 
peaks in the $\sigma^2_{xy}$ spectrum in Fig. \ref{fig:bmloptics}(d) are indicated by red and blue arrows,
respectively. Also, the optical transitions in the vicinity of the $\Gamma$ point are indicated 
by the magenta arrows, namely, A1 is at (0.04,0.04,0) $2\pi/a$ and A2 is at (0.03,0.03,0) $2\pi/a$. } 
\label{fig:b-selec}
\end{figure}

\begin{figure}[t] \centering
\includegraphics[width=8.6cm]{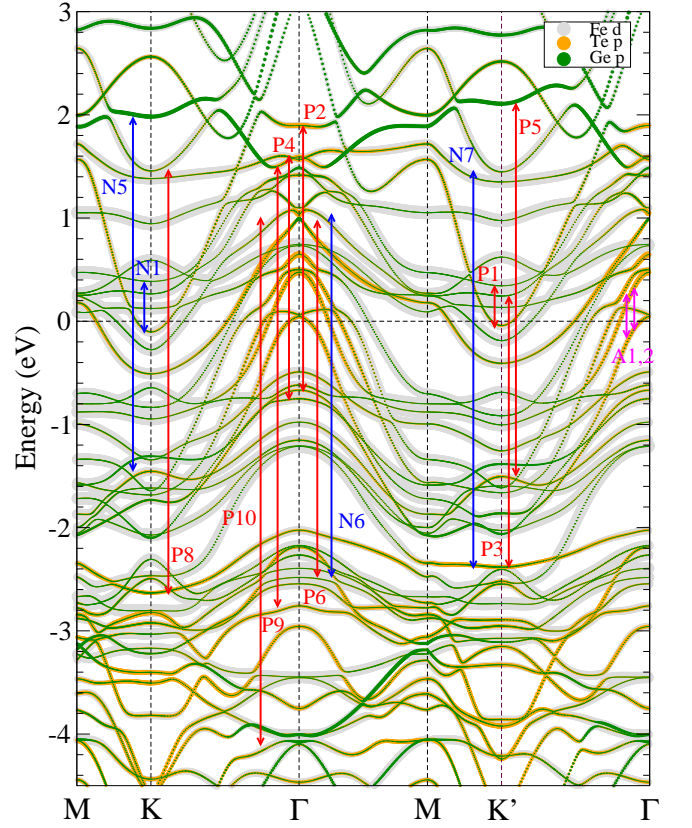}
\caption{Relativistic site- and orbital-projected band structures of ferromagnetic Fe$_3$GeTe$_2$ ML. The 
Fermi level is at 0 eV. The main interband transitions at the $\Gamma$, K and K$'$ point, as well as the 
corresponding peaks in the $\sigma^2_{xy}$ spectrum in Fig. \ref{fig:bmloptics}(h) are indicated by red and blue 
arrows, respectively. Also, the optical transitions in the vicinity of the $\Gamma$ point 
are indicated by the magenta arrows, namely, A1 is at (0.08,-0.04,0) $2\pi/a$ 
and A2 is at (0.07,-0.035,0)$ 2\pi/a$.} 
\label{fig:ml-selec}
\end{figure}

Equations (3), (4) and (11) indicate that the absorptive parts of the optical conductivity elements 
($\sigma^1_{xx}$, $ \sigma^1_{zz}$, $\sigma^2_{xy}$, $\sigma^1_\pm$) are directly related to 
the dipole-allowed interband transitions. This allows us to further understand the origin of the prominent peaks 
in the optical spectra in terms of the band state symmetries and dipole selection rules  
of the considered materials here. To this end, we perform a symmetry analysis on
the band states and dipole selection rules at high symmetry $\Gamma$ and K points,
as described in the supplementary note 1 in the SM~\cite{SM}. The derived dipole selection rules
are listed in Tables S2 and S3, and the calculated optical transition matrix elements 
are given in Tables S4 and S5. 
As mentioned before, few-layer Fe$_3$GeTe$_2$ with an odd number of MLs have the broken inversion symmetry
and thus the K$'$ and K points are not equivalent. Therefore, for ML Fe$_3$GeTe$_2$, we include  the K$'$ point  
in our symmetry analysis as well. Figures \ref{fig:b-selec} and \ref{fig:ml-selec} depict 
the orbital-projected relativistic band structures of bulk and ML Fe$_3$GeTe$_2$, respectively.
We label the main dipole-allowed optical transitions on high symmetry points $\Gamma$ 
and K of bulk Fe$_3$GeTe$_2$ in Fig. \ref{fig:b-selec} and also on high symmetry points $\Gamma$,
K and  K$'$ for ML Fe$_3$GeTe$_2$ in Fig. \ref{fig:ml-selec}. 
Based on the calculated transition matrix elements (Tables S4 and S5) 
and the derived selection rules (Tables S2 and S3), 
we assign the peaks in the absorptive part of the magnetic-optical conductivity $\sigma^2_{xy}$ 
[see Figs. \ref{fig:bmloptics}(d) and \ref{fig:bmloptics}(h)] to the main optical transitions 
near the high symmetry points in Figs.  \ref{fig:b-selec} and \ref{fig:ml-selec}, as indicated 
by red, blue and magenta symbols and arrows. 

We can see from Figs. \ref{fig:b-selec} and \ref{fig:ml-selec} that both bulk and ML Fe$_3$GeTe$_2$ 
have major optical transitions in the energy range from -4.0 eV to 2.0 eV. 
Interestingly, we find that in this energy range, 
significant optical transitions would occur only from Fe $d$ orbital dominated states 
which hybridize with Te $p$ and Ge $p$ orbitals. 
This results from a direct impact of the strong SOC of heavy elements 
on the MO effects \cite{Li2020}. Thus, we could also link the main transitions to 
the calculated DOS spectra presented in the previous section. 
For example, peak P9 of $\sigma^2_{xy}$ of bulk Fe$_3$GeTe$_2$ [Fig. \ref{fig:bmloptics}(d)] 
originates mainly from an optical transition from the valence states at -3.0 eV to 
the conduction states at $\sim$ 2.0 eV [Fig. \ref{fig:b-selec}], which coincides with the Te peak A 
to the Ge peak B in Figs. \ref{fig:bdos}(d) and \ref{fig:bdos}(e), respectively. 
Similarly, peak P10 of bulk Fe$_3$GeTe$_2$ [Fig. \ref{fig:bmloptics}(d)] originates mainly 
from an optical transition from -4.0 eV to $\sim$ 0.8 eV [Fig.  \ref{fig:b-selec}], 
which coincides with the Ge peak C to the FeI and FeII peak D from Figs. \ref{fig:bdos}(b) and \ref{fig:bdos}(e). 
Such connection can also be seen in ML Fe$_3$GeTe$_2$ 
since the DOS spectra from the two structures are similar. 
Furthermore, we show that the considerable hybridization of Fe $d$ orbitals 
with $p$ orbitals of heavy elements in the vicinity of the $\Gamma$ point 
in the energy range from around -0.2 eV to $\sim$0.2 eV also contribute to the first prominent peak 
in all the considered Fe$_3$GeTe$_2$ structures. We indicate two of such transitions as a representative 
by magenta arrows in Figs. \ref{fig:b-selec} and \ref{fig:ml-selec}.  
 
Importantly, the main optical transitions would reveal not only the effect of orbital hybridizations 
but also the impact of crystalline symmetry and the SOC. 
First let us look into the effect of SOC-lifted degeneracies. 
Figures \ref{fig:b-selec} and \ref{fig:ml-selec} show that pairs of right- and left-circular dipole-allowed 
transitions appear, such as peaks P7 and N7 in Fig. \ref{fig:b-selec} for bulk Fe$_3$GeTe$_2$
and also peaks P6 and N6 in Fig. \ref{fig:ml-selec} for ML Fe$_3$GeTe$_2$. 
As explained in the supplementary note 1~\cite{SM}, lifted degeneracies would lead to different 
irreducible representations (irreps) and thus difference in the optical transitions
due to left- and right-circularly polarized light. From Table S4 and S5 we could directly observe 
the magnetic circular dichroism by sign and value differences of the transition matrix elements. 
Second, we examine the effect of the crystalline symmetry. Figure \ref{fig:ml-selec} indicates that 
optical transitions of ML Fe$_3$GeTe$_2$ sometimes come in pairs at the K and K$'$ point, e.g., 
peaks N1 and P1 as well as peaks N5 and P5. This results from the inversion symmetry breaking which 
leads to an interchange of irreps between the K and K$'$ point. With the inclusion of the SOC 
and the spontaneous exchange field due to the intrinsic magnetization, ML Fe$_3$GeTe$_2$ could be 
a possible ferrovalley material \cite{Tong2016}. From Table S5 in the SM~\cite{SM}, we can see the transition 
energy ($\Delta E_{ij}$) to be different for transitions P1 and N1. 
However, in the absence of ferromagnetism, the dipole-allowed transitions would still be 
of opposite chirality but with the same $\Delta E_{ij}$ due to the broken inversion symmetry. 
The large difference of $\sim$0.13 eV in $\Delta E_{ij}$ is thus an indication of possible strong ferrovalley effect
in few-layer Fe$_3$GeTe$_2$ containing odd number of MLs.  
Nevertheless, the $\sigma^2_{xy}$ spectra from different few-layer structures 
(Fig. \ref{fig:bmloptics} as well as Figs. S4 and S5) are similar.
This suggests that although the broken inversion symmetry in the odd-layer structures affects 
the transition matrix elements at the K and K$'$ point, it does not change the magneto-optical spectrum significantly. 
Finally, it should be emphasized that the crucial factor for the large MO conductivity in these structures 
is the orbital hybridization between the magnetic (Fe) atoms which introduce magnetization 
and heavy elements (Te) which bring about strong SOC.~\cite{Guo1996} 

\subsection{Magneto-optical Kerr and Faraday effects}

\begin{figure}[t] \centering
\includegraphics[width=8.6cm]{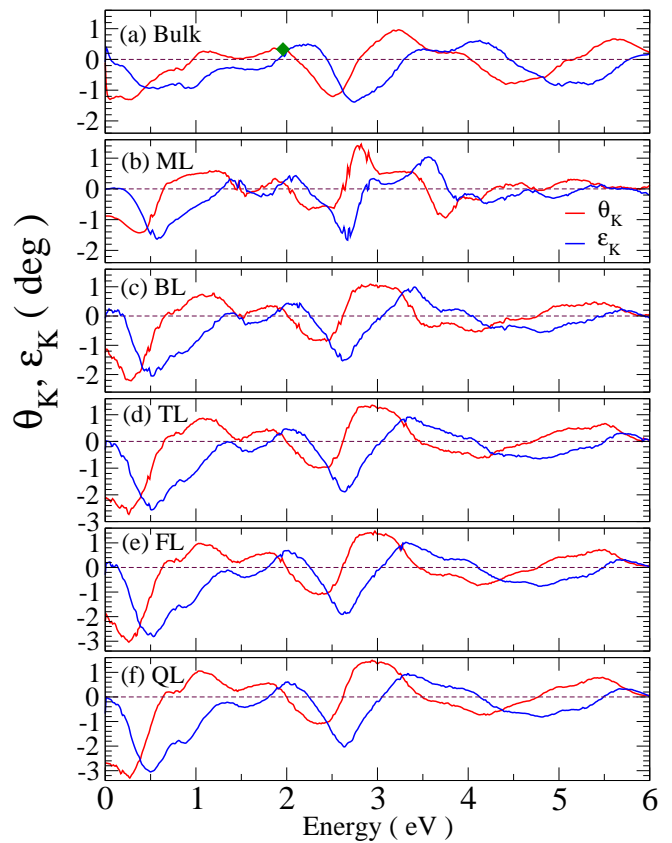}
\caption{Kerr rotation ($\theta_K$) and ellipticity ($\epsilon_K$) spectra for (a) bulk, (b) ML, (c) BL, 
(d) TL, (e) FL and (f) QL Fe$_3$GeTe$_2$ in the ferromagnetic state with the out-of-plane magnetization. 
In (a), the green diamond denotes the experimental $\theta_K$ value~\cite{Fei2018}.} 
\label{fig:kerrot}
\end{figure}

\begin{figure}[t] \centering
\includegraphics[width=8.6cm]{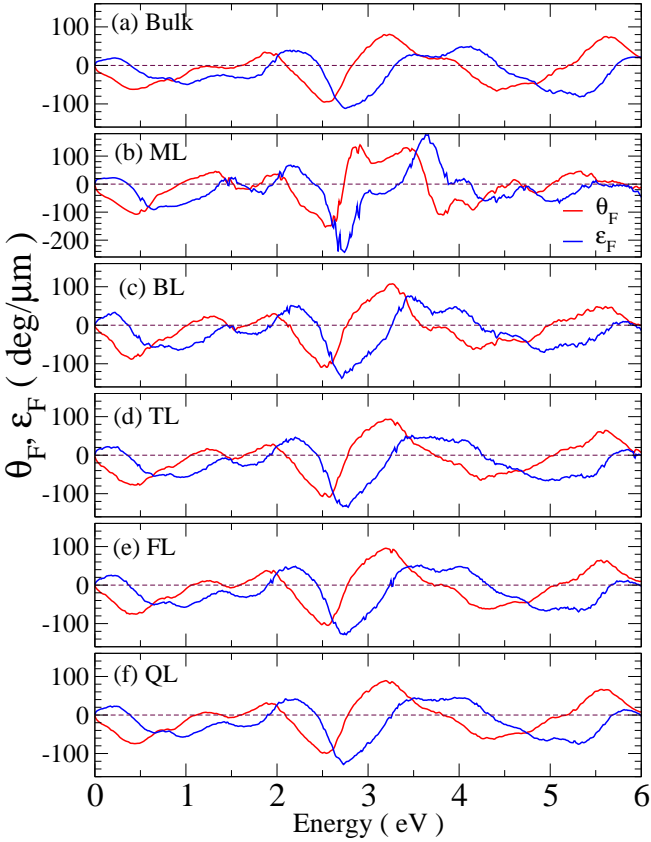}
\caption{Faraday rotation ($\theta_K$) and ellipticity ($\epsilon_K$) spectra for (a) bulk, (b) ML, (c) BL, 
(d) TL, (e) FL and (f) QL Fe$_3$GeTe$_2$ in the ferromagnetic state with the out-of-plane magnetization.} 
\label{fig:fararot}
\end{figure}

Finally, we plot the calculated MOKE and MOFE spectra as a function of photon energy 
in Fig. \ref{fig:kerrot} and \ref{fig:fararot}, respectively. 
Figures \ref{fig:kerrot} and \ref{fig:fararot} show that for all the few-layer structures of Fe$_3$GeTe$_2$, 
the patterns of the MOKE and MOFE spectra look similar to that of 
bulk Fe$_3$GeTe$_2$ [Fig. \ref{fig:kerrot}(a) and Fig. \ref{fig:fararot}(a)], 
especially the Kerr and Faraday rotation spectra of BL, TL, FL and QL Fe$_3$GeTe$_2$. 
As for the optical conductivity spectra, this similarity of the MO spectra 
among all the considered structures is due to the weak interlayer coupling in these Fe$_3$GeTe$_2$ systems.

Figure \ref{fig:kerrot} also shows that bulk and few-layer Fe$_3$GeTe$_2$ all exhibit large negative Kerr 
rotations above 1.0 $^\circ$ in the low energy region below $\sim$0.4 eV. 
In particular, the Kerr angles of FL and QL Fe$_3$GeTe$_2$ are larger than 3.0 $^\circ$ at $\sim$0.3 eV.
As photon energy further increases, the negative Kerr rotations decrease monotonically
and become positive between 0.6 eV and 1.0 eV. Then they all peak at around 1.0 eV and
stay positive until photon energy of around 2.0 eV. The positive peaks near 1.0 eV
are rather prominent and can reach 1.0 $^\circ$ in TL, FL and QL Fe$_3$GeTe$_2$ (Fig. \ref{fig:kerrot}). 
The Kerr rotation spectra then have a negative peak of -1.0 $^\circ$ again at around 2.5 eV
and become positive again as photon energy increases beyond $\sim$2.6 eV.
They all have a positive pronounced peak at around 3.0 eV except bulk and ML Fe$_3$GeTe$_2$
which have the positive peak at 3.2 eV and 2.8 eV, respectively. 
Interestingly, the Kerr ellipticity spectra of all the structures have two large negative peaks
at around 0.5 eV and 2.6 eV, respectively. 
MOKE experiments were carried out on a 340-nm thick sample using 633 nm HeNe laser (photon energy 1.96 eV).
We note that the measured Kerr angle of $\sim$0.33$^\circ$ \cite{Fei2018} is compared well
with our calculated Kerr angle of bulk Fe$_3$GeTe$_2$ [see Fig. \ref{fig:kerrot}(a)]. 

To investigate the potential applications of the Fe$_3$GeTe$_2$ systems in, e.g., MO devices, 
we now  compare the magnitudes of Kerr rotation angles to several popular MO materials. 
Let us start with 3$d$ transition metals and their alloys.
We notice that bcc Fe metal has a Kerr rotation of -0.5$^\circ$ at 1.0 eV, hcp Co metal 
has a Kerr rotation of -0.42$^\circ$ at 5.0 eV, and fcc Ni metal has a Kerr rotation of -0.25$^\circ$ 
at 4.0 eV \cite{MO_Antonov_book,Guo1995,Oppeneer1992}. Clearly, the calculated Kerr rotation angles
of few-layer and also bulk Fe$_3$GeTe$_2$ are generally larger than that of elemental 3$d$ transition metals.
In particular, the magnitudes of both the Kerr rotation angle and ellipticity of
all the considered Fe$_3$GeTe$_2$ structures are larger
than 1.0$^\circ$ at around 2.5 eV.
They are also comparable or even larger than the MOKE in 3$d$ transition metal alloys with heavy
elements such as Pt and Bi which have the strong SOC.~\cite{Guo1996}
For example, famous MO transition metal alloys such as FePt, CoPt and PtMnSb 
have Kerr rotation angles ranging from 0.4$^\circ$ to 0.5$^\circ$ \cite{Guo1995,Engen1998}. 
Nevertheless, thin film MnBi has a record-high Kerr rotation 
of 2.3$^\circ$ at 1.84 eV \cite{Di1996}, which is comparable to or smaller than that
of TL, FL and QL Fe$_3$GeTe$_2$ in the infra-red frequency range (within 1.0 eV)
(see Fig. \ref{fig:kerrot}). 

For further comparison, let us examine famous magnetic semiconductors with good MO properties. Ferromagnetic
semiconductors Y$_3$Fe$_5$O$_{12}$ and Bi$_3$Fe$_5$O$_{12}$, which have been widely used in spintronic research,
have Kerr rotation angle of -0.12$^\circ$ at 4.8 eV and -1.21$^ \circ$ at 2.4 eV, respectively \cite{Tomita2006,Li2020}. 
For diluted magnetic semiconductors Ga$_{1-x}$Mn$_x$As, a Kerr rotation of 
$\sim$0.4$^\circ$ near 1.8 eV was reported \cite{Lang2005}. 
Excitingly, layered ferromagnetic semiconductors CrI$_3$ and Cr$_2$Ge$_2$Te$_6$ were 
recently thinned down to just one or two monolayer with the ferromagnetic order retained at low 
temperatures~\cite{Huang2017,Gong2017}. Moreover, TL Cr$_2$Ge$_2$Te$_6$ and TL CrI$_3$ were predicted 
to have large Kerr rotation of 0.7$^\circ$ near 2.8 eV and  $\sim$1.0$^\circ$ at 1.3 eV, respectively 
\cite{Yimei2018,Vijay2019}. Overall, in comparison to these famous MO semiconductors, the MOKE in  
bulk and ML Fe$_3$GeTe$_2$ are comparable and in particular, TL, FL and QL Fe$_3$GeTe$_2$ 
exhibit stronger MOKE. 

Remarkably, from Fig. \ref{fig:fararot}, we can observe large positive (negative) peaks of value 82 
(-97) deg/$\mu m$, 142 (-156) deg/$\mu m$, 109 (-111) deg/$\mu m$, and 95 (-111) deg/$\mu m$ at 
$\sim$2.55 (3.0) eV for bulk, ML, BL, and TL Fe$_3$GeTe$_2$, respectively. For comparison, MnBi thin films are known to 
possess the largest Faraday rotation angle of $\sim$80 deg/$\mu m$ at 1.77 eV~\cite{Ravindran1999,Di1996}. 
The widely used semiconductor Y$_3$Fe$_5$O$_{12}$ possesses a 
Faraday rotation of 7.2 deg/$\mu m$ at 3.9 eV \cite{Li2020}. With the substitution of Y by
heavy element bismuth, Bi$_3$Fe$_5$O$_{12}$ can show a large Faraday 
rotation angle of 51.2 deg/$\mu m$ at 3.7 eV~\cite{Li2020}. 
We notice that recently discovered 2D ferromagnetic semiconductors Cr$_2$Ge$_2$Te$_6$ 
and CrI$_3$ have Faraday rotation angles of $\sim$120 deg/$\mu m$~\cite{Yimei2018}
and of at most $\sim$ 108 deg/$\mu m$~\cite{Vijay2019,Wu2019}, respectively. 
Clearly, bulk and 2D Fe$_3$GeTe$_2$ reported here have relatively large Faraday rotation angles. 
The outstanding MO properties of the considered Fe$_3$GeTe$_2$ systems suggest their promising 
applications for nanoscale MO sensors and high density MO data-storage devices.

Finally, we notice that MO Faraday rotation spectra of bulk and 2D Fe$_3$GeTe$_2$ (Fig. \ref{fig:fararot})
are similar to their MO Kerr rotation spectra (Fig. \ref{fig:kerrot}).
This may be expected since Eqs. (7), (8) and (10) indicate that all these MO spectra
are more or less proportional to the MO conductivity ($\sigma_{xy}$).
Indeed, the large MOKE and MOFE in bulk and few-layer Fe$_3$GeTe$_2$ stem from
their large MO conductivity (i.e., strong magnetic circular dichroism).
For example, overall, the MO conductivity of bulk and 2D Fe$_3$GeTe$_2$ 
(Fig. 5 as well as Figs. S4 and S5 in the SM~\cite{SM}) is 
about 20 times larger than Y$_3$Fe$_5$O$_{12}$ and also 2 times larger than Bi$_3$Fe$_5$O$_{12}$~\cite{Li2020}.
They are also around 2 times larger than bulk and 2D CrI$_3$~\cite{Vijay2019}.

\section{CONCLUSIONS}

Summarizing, we have investigated two relativity-induced properties, namely, magnetic anisotropy energy (MAE) and
magneto-optical (MO) effects, of ML, BL, TL, FL and QL as well as bulk Fe$_3$GeTe$_2$, 
and also their connections with the underlying electronic structures of the materials, 
based on systematic first principle DFT calculations.
First of all, we find that all the considered Fe$_3$GeTe$_2$ structures prefer the out-of-plane magnetization
and have gigantic MAEs of $\sim$3.0 meV/f.u., being about 20 and 6 times larger than
2D ferromagnetic semiconductors Cr$_2$Ge$_2$Te$_6$ and CrI$_3$, respectively.
These MAE values are also comparable to that of FePt, which is known to have the
largest MAE among the magnetic transition metals and their alloys.
This gigantic perpendicular anisotropy results from the large relativity-induced MCE of $\sim$3.32 meV/f.u.,
which is ten times larger than the competing classical MDE of $\sim$0.3 meV/f.u., which favors
an in-plane magnetization (Table II). The giant MCEs can be attributed to the presence of 
significant Te $p_{x,y}$ orbital DOS in the vicinity the Fermi level and also to some extent, 
to the topological nodal point just below the Fermi level at the K points in the Brillouin zone 
in these materials. This strong PMA thus stabilizes the long-range ferromagnetic order
in 2D Fe$_3$GeTe$_2$ at temperatures higher than that in 2D  Cr$_2$Ge$_2$Te$_6$ and CrI$_3$. 

Secondly, we also find that 2D and bulk Fe$_3$GeTe$_2$ exhibit strong MO effects 
with the calculated Kerr and Faraday rotation angles being comparable or even larger 
than that of best-known bulk MO materials such as PtMnSb, Y$_3$Fe$_5$O$_{12}$ and Bi$_3$Fe$_5$O$_{12}$. 
In particular, all the Fe$_3$GeTe$_2$ structure are predicted to have large Kerr rotation angles
of $\sim$1.0$^\circ$ at $\sim$3.0 eV and FL and QL Fe$_3$GeTe$_2$ have Kerr rotation angles 
as large as $\sim$3.0$^\circ$ at $\sim$0.25 eV (Fig. 8).
Furthermore, ML Fe$_3$GeTe$_2$ has a Faraday rotation angle of -156 deg/$\mu$m, which is three times larger than
that of famous MO oxide Bi$_3$Fe$_5$O$_{12}$.
The shape and position of the features in the Kerr and Faraday rotation spectra
are almost thickness-independent though the Kerr rotation angles increase slightly with the film thickness.
The strong MO Kerr and Faraday effects are found to result from
the large MO conductivity (i.e., strong magnetic circular dichroism) in these ferromagnetic materials.
In particular, the calculated MO conductivity spectra are one order of magnitude larger than that of Y$_3$Fe$_5$O$_{12}$.
The calculated MO conductivity spectra are analysed in terms of the symmetry of the band states
and dipole-allowed optical transitions at high symmetry $\Gamma$, K and K$'$ points,
which also reveal that atomically thin Fe$_3$GeTe$_2$ films with odd layer-number
would exhibit anomalous ferrovalley Hall effect.
We notice that our calculated MAE values for bulk and ML Fe$_3$GeTe$_2$ agree well with the
corresponding experimental values and our predicted Kerr angle of bulk Fe$_3$GeTe$_2$ at 1.96 eV
is in good agreement with the measured one. Also, the DC-limit of the real part of the calculated 
MO conductivity of bulk Fe$_3$GeTe$_2$ agrees well with the measured AHC. 
ALL these interesting findings therefore indicate that 2D and bulk ferromagnetic Fe$_3$GeTe$_2$
may find valuable applications for high density MO and spintronic nanodevices.

\section*{ACKNOWLEDGMENTS}
The authors acknowledge the support from the Ministry of Science and Technology and the 
National Center for Theoretical Sciences (NCTS) of The R.O.C. The authors are also grateful to the National
Center for High-performance Computing (NCHC) for the computing time.


\begin{thebibliography}{}

\bibitem{Gong2017} C. Gong, L. Li, Z. Li, H. Ji, A. Stern, Y. Xia, T. Cao, W. Bao, C. Wang, Y. Wang, Z. Q. 
Qiu, R. J. Cava, S. G. Louie, J. Xia, and X. Zhang, Discovery of intrinsic ferromagnetism in two-dimensional van der 
Waals crystals, Nature \textbf{546}, 265 (2017).

\bibitem{Huang2017} B. Huang, G. Clark, E. Navarro-Moratalla, D. R. Klein, R. Cheng, K. L. Seyler, D. Zhong, E.
Schmidgal, M. A. McGuire, D. H. Cobden, W. Yao, D. Xiao, P. Jarillo-Herrero, and X. Xu, Layer-dependent
ferromagnetism in a van der Waals crystal down to monolayer limit, Nature \textbf{546}, 270 (2017).

\bibitem{Fei2018} Z. Fei1, B. Huang, P. Malinowski1, W. Wang, T. Song, J. Sanchez1, W. Yao, D. Xiao, X. Zhu,
A. F. May, W. Wu,D. H. Cobden, J.-H. Chu, and X. Xu, Two-dimensional itinerant ferromagnetism atomically thin Fe
$_3$GeTe$_2$, Nat. Mat. \textbf{17}, 778 (2018).

\bibitem{Deng2018} Y. Deng, Y. Yu1, Y. Song, J. Zhang, N. Z. Wang, Z. Sun, Y. Yi,Y. Z. Wu, S. Wu, J. Zhu, J. 
Wang, X. H. Chen, and Y. Zhang, Gate-tunable room-temperature ferromagnetism in two-dimensional Fe$_3$GeTe$_2$, 
Nature \textbf{563}, 94 (2018).

\bibitem{Mermin1966} N. D. Mermin and H. Wagner, Absence of ferromagnetism or antiferromagnetism in one- or two-
dimensional isotropic Heisenberg models, Phys. Rev. Lett. \textbf{17}, 1133 (1966).

\bibitem{Gibertini2019} M. Gibertini, M. Koperski, A. F. Morpurgo, and K. S. Novoselov, Magnetic 2D materials 
and heterostructures, Nat. Nanotechnol. \textbf{14}, 408 (2019).

\bibitem{Burch2018} K. S. Burch, D. Mandrus, and J.G. Park, Magnetism in two-dimensional van der Waals 
material, Nature \textbf{563}, 47 (2018).

\bibitem{Zhuang2016} H. L. Zhuang, P. R. C. Kent, and R. G. Hennig, Strong anisotropy and magnetostriction
in the two-dimensional Stoner ferromagnet Fe$_3$GeTe$_2$, Phys. Rev. B \textbf{93}, 134407 (2016).

\bibitem{Wang2020} Y.-P. Wang, X.-Y. Chen, and M.-Q. Long, Modifications of magnetic anisotropy of Fe
$_3$GeTe$_2$ by the electric field effect, Appl. Phys. Lett. \textbf{116}, 092404 (2020).

\bibitem{Lin2019} X. Lin and J. Ni, Layer-dependent intrinsic anomalous Hall effect in Fe$_3$GeTe$_2$,
Phys. Rev. B \textbf{100}, 085403 (2019).

\bibitem{Tong2016} W.-Y. Tong, S.-J. Gong, X. Wan, C.-G. Duan, Concepts of ferrovalley material and anomalous valley
Hall effect, Nat. Comms \textbf{7}, 13612 (2016).

\bibitem{Gudelli2020} V. K. Gudelli and G.-Y. Guo, Antiferromagnetism-induced second-order nonlinear optical
responses of centrosymmetric bilayer CrI$_3$, Chin. J. Phys. \textbf{68}, 896 (2020).

\bibitem{Guo1991A} G.-Y. Guo, W. M. Temmerman, and H. ebert, First-principles determination of the
magnetization direction of Fe monolayer in noble metals, J. Phys.: Condens. Matter \textbf{3}, 8205 (1991).

\bibitem{Guo2007} J. C. Tung, and G.-Y. Guo, Systematic \textit{ab initio} study of the magnetic and
electronic properties of all 3$d$ transition metal linear and zigzag nanowires, Phys. Rev. B \textbf{76}, 094413
(2007).

\bibitem{Yimei2018} Y. Feng, S. Wu, Z.-Z. Zhu, and G.-Y. Guo, Large magneto-optical effects and magnetic
anisotropy energy in two-dimensional Cr$_2$Ge$_2$Te$_6$, Phys. Rev. B \textbf{98}, 125416 (2018).

\bibitem{Park2020} S. Y. Park, D. S. Kim, Y. Lin, J. Hwang, Y. Kim, W. Kim, J.-Y. Kim, C. Petrovic, C.
Hwang, S.-K. Mo, H.-j. Kim, B.-C. Min, H. C. Koo, J. Chang, C. Jang, J. W. Choi, and H. Ryu, Controlling the Magnetic
Anisotropy of the van der Waals Ferromagnet Fe$_3$GeTe$_2$ through Hole Doping, Nano Lett. \textbf{20}, 95, (2020).

\bibitem{MO_Oppeneer_book} P. M. Oppeneer, Chapter 1 Magneto-optical Kerr Spectra, pp. 229-422, in Handbook of
Magnetic Materials, edited by K. H. J. Buschow. Elsevier, Amsterdam, (2001).

\bibitem{MO_Antonov_book} V. Antonov, B. Harmon, and A. Yaresko, Electronic structure and magneto-optical properties
of solids, Springer Science \& Business Media, (2004).

\bibitem{MO_storage} M. Mansuripur, The Principles of Magneto-Optical Recording, (Cambridge University Press,
Cambridge, 1995).

\bibitem{MO_device} J. P. Castera, in Magneto-optical Devices, Vol. 9 of Encyclopedia of Applied Physics, edited by
G. L. Trigg (Wiley-VCH, New York, 1996), p. 133.

\bibitem{Guo1995} G.-Y. Guo, H. Ebert, Band-theoretical investigation of the magneto-optical Kerr
effect in Fe and Co mulitlayers, Phys. Rev. B \textbf{51}, 12633 (1995).

\bibitem{Guo1996} G.-Y. Guo, H. Ebert, On the origins of the enhanced magneto-optical Kerr effect in
ultrathin Fe and Co multilayers, J. Magn. Magn. Mater. \textbf{156}, 173 (1996).

\bibitem{Deiseroth2006} H.-J. Deiseroth, K. Aleksandrov, C. Reiner, L. Kienle, and R. K. Kremer, Fe$_3$GeTe
$_2$ and Ni$_3$GeTe$_2$ – Two New Layered Transition-Metal Compounds: Crystal Structures, HRTEM Investigations, and
Magnetic and Electrical Properties, Eur. J. Inorg. Chem. \textbf{2006}, 1561 (2006).

\bibitem{Perdew1996} J. P. Perdew, K. Burke, and M. Ernzerhof, Generalized Gradient Approximation Made Simple,
Phys. Rev. Lett. \textbf{77}, 3865 (1996).

\bibitem{Grimme2006} S. Grimme, Semiempirical GGA-Type Density Functional Constructed
with a Long-Range Dispersion Correction, J. Comput. Chem. \textbf{27}, 1787 (2006).

\bibitem{Blochl1994} P. E. Blochl, Projector augmented-wave method, Phys. Rev. B \textbf{50}, 17953 (1994).

\bibitem{Kresse1993} G. Kresse and J. Hafner, \textit{Ab initio} molecular dynamics for liquid metals, Phys.
Rev. B \textbf{47}, 558 (1993).

\bibitem{Kresse1996} G. Kresse and J. Furthmuller, Efficient interactive schemes for \textit{ab initio} total-
energy calculations using a plane-wave basis set, Phys. Rev. B \textbf{54}, 11169 (1996).

\bibitem{Jepson1971} O. Jepson and O. K. Anderson, The electronic structure of h.c.p Ytterbium, Solid State
Commun. \textbf{9}, 1763 (1971).

\bibitem{Vijay2019} V. K. Gudelli, and G.-Y. Guo, Magnetism and magneto-optical effects in bulk and few-
layer CrI$_3$ : a theoretical GGA + $U$ study, New J. Phys. \textbf{21}, 053012 (2019).

\bibitem{Wang1974} C. S. Wang and J. Callaway, Band structure of nickel: Spin-orbit coupling, the Fermi 
surface, and the optical conductivity, Phys. Rev. B \textbf{9}, 4897 (1974).

\bibitem{Oppeneer1992} P. M. Oppeneer, T. Maurer, J. Sticht, and J. K$\ddot{u}$bler, \textit{An initio} 
calculated magneto-optical Kerr effect of ferromagnetic metals: Fe and Ni, Phys. Rev. B \textbf{45}, 10924 (1992).

\bibitem{Feng2015} W. Feng, G.-Y. Guo, J. Zhou, Y. Yao, and Q. Niu, Large magneto-optical Kerr effect in 
noncollinear antiferromagnets Mn$_3$ $X$ ($X$=Rh, Ir, Pt), Phys. Rev. B \textbf{92}, 144426 (2015).


\bibitem{Adolph2001} B. Adolph, J. Furthmuller, and F. Bechstedt, Optical properties of semiconductors using 
projector-augmented waves, Phys. Rev. B \textbf{63}, 125108 (2001).

\bibitem{Temmerman1989} W. M. Temmerman, P. A. Sterne, G.-Y. Guo, and Z. Szotek, Electronic Structure 
Calculations of High T$_c$ Materials, Mol. Simul. \textbf{63}, 153 (1989).

\bibitem{Guo1994} G.-Y. Guo, and H. Ebert, Theoretical investigation of the orientation dependence of the 
magneto-optical Kerr effect in Co, Phys. Rev. B \textbf{50}, 10377(R), 1994.

\bibitem{Sivadas2016} N. Sivadas, S. Okamoto, and D. Xiao, Gate-Controllable Magneto-optic Kerr Effect 
in Layered Collinear Antiferromagnets, Phys. Rev. Lett. \textbf{117}, 267203 (2016).

\bibitem{Wu2019} M. Wu, Z. Li, T. Cao and S. G. Louie, Physical origin of giant excitonic 
and magnetooptical responses in two-dimensional ferromagnetic insulators, 
Nature Commun. \textbf{10}, 2371 (2019).

\bibitem{Ravindran1999} P. Ravindran, A. Delin, P. James, B. Johansson, J. M. Wills, R. Ahuja, and O. 
Eriksson, Phys. Rev. B \textbf{59}, 15680 (1999). 

\bibitem{Chen2013} B. Chen, J.H. Yang, H.D. Wang, M. Imai, H. Ohta, C. Michioka, K. Yoshimura, and M.H. Fang, 
Magnetic Properties of Layered Itinerant Electron Ferromagnet Fe$_3$GeTe$_2$, J. Phys. Soc. Jpn. \textbf{82}, 124711 
(2013).

\bibitem{May2016} A. F. May, S. Calder, C. Cantoni, H. Cao, and M. A. McGuire, Magnetic structure and 
phase stability of the van der Waals bonded ferromagnet Fe$_{3-x}$GeTe$_2$, Phys. Rev. B \textbf{93}, 014411 (2016).

\bibitem{Zhu2016} J.-X. Zhu, M. Janoschek, D. S. Chaves, J. C. Cezar, T. Durakiewicz, F. Ronning, Y. Sassa, 
M. Mansson, B. L. Scott, N. Wakeham, E. D. Bauer, and J. D. Thompson, Electronic correlation and magnetism in the 
ferromagnetic metal Fe$_3$GeTe$_2$, Phys. Rev. B \textbf{93}, 144404 (2016).

\bibitem{Verchenko2015} V. Y. Verchenko, A. A. Tsirlin, A. V. Sobolev, I. A. Presniakov, and A. V. Shevelkov,
Ferromagnetic Order, Strong Magnetocrystalline Anisotropy, and Magnetocaloric Effect in the Layered Telluride Fe$_{3-
\delta}$GeTe$_2$, Inorg. Chem. \textbf{54}, 8598 (2015).

\bibitem{Guo1997} G. Y. Guo, Spin and Orbital Polarized Multiple Scattering Theory
of Magneto-x-ray Effects in Fe, Co and Ni, Phys. Rev. B \textbf{55}, 11619 (1997).

\bibitem{Oppeneer1998} P. M. Oppenner, Magneto-optical spectroscopy in the valence-band energy regime:
relationship to the magnetocrystalline anisotropy, J. Magn. Magn. Mater. \textbf{188}, 275 (1998).

\bibitem{Guo1991B} G.-Y. Guo, W. M. Temmerman, and H. Ebert, A relativistic spin-polarized band theoretical
study of magnetic properties of nickel and iron, Physica B Condens. Matter \textbf{172}, 61 (1991).

\bibitem{Bruno1989} P. Bruno, Tight-binding approach to the orbital magnetic moment and magnetocrystalline
anisotropy of transition-metal monolayers, Phys. Rev. B \textbf{39}, 865 (1989).

\bibitem{SM} See Supplemental Material at http://link.aps.org/supplemental/ for Figs. S1-S5,
supplementary note 1 and Tables S1-S5, which include Refs. ~\cite{Kim2018,dresselhaus,koster,Newmarch1982,Newmarch1983,Gao2020}.

\bibitem{Kim2018} K. Kim, J. Seo, E. Lee, K.-T. Ko, B. S. Kim, B. G. Jang, J. M. Ok, J. Lee, Y. J. Jo,
W. Kang, J. H. Shim, C. Kim, H. W. Yeom, B. I. Min, B.-J. Yang , and J. S. Kim, Large anomalous Hall current induced
by topological nodal lines in a ferromagnetic van der Waals semimetal, Nat. Mater. \textbf{17}, 794 (2018).

\bibitem{dresselhaus}  M. S. Dresselhaus, G. Dresselhaus, and A. Jorio, \textit{Group Theory Applications to the 
Physics of Condensed Matter} (Springer-Verlag, Berlin, 2008).

\bibitem{koster} G. F. Koster, J. O. Dimmock, R. G. Wheeler, and H. Statz, \textit{Properties of the thirty-two point 
groups} (Cambridge, Mass.: M.I.T. Press., 1963).

\bibitem{Newmarch1982} J. D. Newmarch, and R. M. Golding, The character table for the corepresentations of magnetic
groups, J. Math. Phys. \textbf{23}, 695 (1982).

\bibitem{Newmarch1983} J. D. Newmarch, Some character theory for groups of linear and antilinear operators, J. Math.
Phys. \textbf{24}, 742 (1983).

\bibitem{Gao2020} J. Gao, Q. Wu, C. Persson and Z. Wang, Irvsp: to obtain irreducible representations of electronic
states in the VASP, Comput. Phys. Commun. \textbf{261}, 107760 (2021).

\bibitem{Wang1993} D.-s. Wang, R. Wu, and A. J. Freeman, First-principles theory of surface
magnetocrystalline anisotropy and the diatomic-pair model, Phys. Rev. B \textbf{47}, 14932 (1993).

\bibitem{Takayama1976} H. Takayama, K.-P. Bohnen, and P. Fulde, Magnetic surface anisotropy of transition
metals, Phys. Rev. B \textbf{14}, 2287 (1976).

\bibitem{Xu2019} J. Xu, W. A. Phelan, and C.-L. Chien, Large Anomalous Nernst Effect in a van der Waals
Ferromagnet Fe$_3$GeTe$_2$, Nano Lett. \textbf{19}, 8250 (2019).

\bibitem{Wang2019} X. Wang, Z. Li, M. Zhang, T. Hou, J. Zhao, L. Li, A. Rahman, Z. Xu, J. Gong, Z.
Chi, R. Dai, Z. Wang, Z. Qiao, and Z. Zhang, Pressure-induced modification of the anomalous Hall effect in layered Fe
$_3$GeTe$_2$, Phys. Rev. B \textbf{100}, 014407 (2019).

\bibitem{Li2020} W.-K. Li, and G.-Y. Guo, A First Principle Study on Magneto-Optical Effects and Magnetism
in Ferromagnetic Semiconductors Y$_3$Fe$_5$O$_{12}$ and Bi$_3$Fe$_5$O$_{12}$, Phys. Rev. B \textbf{103}, 014439 (2021).

\bibitem{Engen1998} P. G. van Engen, K. H. J. Buschow, R. Jongebreur, and M. Erman, PtMnSb, a material
with very high magneto-optical Kerr effect, Appl. Phys. Lett. \textbf{42}, 202 (1983).

\bibitem{Tomita2006} S. Tomita, T. Kato, S. Tsunashima, S. Iwata, M. Fujii, and S. Hayashi, Magneto-Optical
Kerr Effects of Yttrium-Iron Garnet Thin Films Incorporating Gold Nanoparticles, Phys. Rev. Lett. \textbf{96}, 167402
(2006).

\bibitem{Lang2005} R. Lang, A. Winter, H. Pascher, H. Krenn, X. Liu, and J. K. Furdyna, Polar Kerr effect
studies of Ga$_{1-x}$Mn$_x$As epitaxial films, Phys. Rev. B \textbf{72}, 024430 (2005).

\bibitem{Di1996} G. Q. Di, and S. Uchiyama, Optical and magneto-optical properties of MnBi film, Phys. Rev. B
\textbf{53}, 3327 (1996).

\end{thebibliography}
\end{document}